\documentstyle[12pt,fleqn]{article}
\renewcommand{\theequation}{\arabic{section}.\arabic{equation}}
\begin{document}
\baselineskip=24pt
\pagestyle{plain}
\title{Effective actions of local composite operators ---
case of $\varphi^4$ theory, itinerant electron model, and QED}
\author{K.\ Okumura \and {\em
Department of Physics, Faculty of Science and Technology,}
\\ {\em Keio University, Yokohama 223, Japan}}
\date{April 14, 1994}
\maketitle
\begin{abstract}
The effective action
$\Gamma[\phi]$ is defined from the generating
function(al) $W[J]$ through Legendre transformation and
plays the role of action functional in the zero-temperature
field theory and of generalized thermodynamical function(al) in
equilibrium statistical physics.
A compact graph rule for the effective action $\Gamma[\phi]$ of
a local composite operator is given in this paper.
This long-standing problem of obtaining
$\Gamma[\phi]$ in this case is solved directly
without using the auxiliary field.
The rule is first deduced
with help of
the inversion method, which is a technique for making
the Legendre transformation perturbatively.
It is then proved by using a topological
relation and also by the sum-up rule.
The latter is a technique for making the
Legendre transformation in a graphical language.
In the course of proof
a special role is played by $J^{(0)}[\phi]$,
which is a function(al) of the variable
$\phi$ and defined through the lowest inversion formula.
Here $J^{(0)}[\phi]$ has the meaning of the source
$J$ for the noninteracting system expressed by $\phi$.
Explicitly derived are
the rules for the effective action of
$\langle \varphi(x)^2 \rangle$
in the $\varphi^4$ theory,
of the number density
$\langle n_{{\bf r}\sigma} \rangle$ in the itinerant
electron model, and of the gauge invariant operator
$\langle \bar{\psi}\gamma^\mu\psi \rangle$ in QED.
\end{abstract}
\maketitle
\newpage
\section{Introduction}
\label{s1}
The effective action $\Gamma[\phi]$ or thermodynamical
function introduced by Legendre transformation
is a very convenient tool in various fields of physics.
Actually this fact has long been realized in the condensed matter
physics as well as in particle physics\cite{REV}.

In spite of its wide-spread use, the precise rule
of constructing the effective
action for a {\em local composite field} has not
been derived up to now although the graphical rules for an
elementary field and for non-local composite fields
up to four-body operators are already known\cite{DM,JL,VKP,J,CJT}.
The study of the effective action for local
composite operator amounts to rewrite the
theory in terms of physical variables such as the
expectation values of the number density operator, spin density
operator,
local gauge invariant operator etc.
Thus the importance of the investigation can not be overestimated.
In the following we deal with three examples ---
the effective action of the $\varphi(x)^2$ operator
in the $\varphi^4$ theory, a generalized free energy
as a function(al) the  spin and
number density in the itinerant electron model,
and the effective action of the  $\bar{\psi}(x)\gamma^\mu\psi(x)$
operator in QED where $\psi$ is the electron field.

In some cases the hard problem to obtain the
effective action of local composite operators
has been avoided
by Hubbard-Stratonovich transformation\cite{HS}
or by introducing an auxiliary field \cite{GN}.
In such a formulation, the auxiliary field is not equal to the
local composite operator if one deals with the off-shell
quantities and extra work is needed to
extract the physical on-shell quantities, which are directly
related to the original local composite operator.
In the following we take the local composite operator itself
without using auxiliary field
and explicitly derive the graphical rule  for
the effective action.
Difficulty is solved by use of
the inversion method \cite{F,MLQED,SUP,UKI,IEM}.

For later discussion let us define the effective action
$\Gamma[\phi]$ explicitly.
For the zero-temperature case
it is introduced through a generating
functional $W[J]$ with a source $J$ coupled to some
operator $\hat{O}$;
$e^{iW[J]}=\langle 0| e^{iJ\hat{O}}|0 \rangle$.
Here $|0\rangle$ represents the ground state.
Then a dynamical variable $\phi$ is defined as
$\phi=\frac{\delta W}{\delta J}\equiv \langle \hat{O} \rangle^J$
and the effective action, which is a functional of $\phi$, is
give by
$\Gamma[\phi]=W[J]-J\phi$
with
$-J=\frac{\delta \Gamma}{\delta \phi}$.
Here
$J$ {\em is given by a functional of} $\phi$
by inverting
$\phi=\frac{\delta W}{\delta J}$.
For simplicity we have considered the $x$-independent
variables $J$
and $\phi$ since it is straightforward to extend to
the local variables $J(x)$ and $\phi(x)$.
We have called a function of $J$ or $\phi$
{\em functional} as we will do in what follows
so that we can recover the $x$-dependence freely.
In equilibrium
statistical physics $W$ corresponds to the thermodynamical
potential $\Omega$ and $\Gamma$ to the Helmholtz free energy
$F$.
For instance, $\hat{O}$ is
chosen to be the total number operator $N$ then
$J$ is the chemical potential $\mu(N)$.

The essential step of Legendre transformation is to
invert the relation $\phi=\frac{\delta W}{\delta J}$.
The inversion method enables us to write down  the explicit form
of $J$ in terms of $\phi$ by perturbative calculation.
The lowest relation of the method defines the functional
$J^{(0)}[\phi]$, which is
the source as a functional of $\phi$ in noninteracting system.
As will become clear, it is $J^{(0)}[\phi]$ that plays
a fundamental role in deriving $\Gamma[\phi]$.
In fact it turns out that
by use of the inversion method $\Gamma[\phi]$ in the case of a local
composite field is obtained as a class
of {\em irreducible} graph in certain sense
(plus simple terms)
as a functional of $J^{(0)}[\phi]$ rather than $\phi$
(for the $\varphi^4$ theory,
see (\ref{GAMa}) with (\ref{J}) or (\ref{GAMc}) below).
In other words, all the functional dependence on $\phi$ is through
$J^{(0)}[\phi]$.
This point is in remarkable contrast to the rules for the effective
action of  an elementary field and non-local composite operators
where the rule is based on $\phi$ itself.
This is the reason why the problem of local composite operator
is difficult and has been unresolved.
The use of $J^{(0)}[\phi]$ naturally comes out in the
formulation through the inversion method.

In order to explain the inversion
method\cite{F,MLQED,SUP,UKI,IEM} again for the simple case of
the $x$-independent variable $J$ and $\phi$, we
assume that the theory has a coupling constant $\lambda$.
Then the expectation value
$\phi =\langle \hat{O} \rangle^J$ is calculated in the
presence of $J$ through
the Feynman rule (like (\ref{PHI1}) below) to get a series expansion
\begin{equation}
\phi=\sum_{n=0}^{\infty}\phi^{(n)}[J]
\label{1.1}
\end{equation}
where $\phi^{(n)}[J]$ is the $n$-th order of $\lambda$
by regarding $J$ as independent of $\lambda$.
This relation can be inverted to give
\begin{equation}
J=\sum_{n=0}^{\infty} J^{(n)}[\phi]
\label{1.2}
\end{equation}
where $J^{(n)}[\phi]$ is the $n$-th order of $\lambda$.
To obtain the explicit form of $J^{(i)}[\phi]$ as a
functional of $\phi$  we first assume (\ref{1.2}) and get
\begin{eqnarray}
\phi&=&
\sum_{n=0}^{\infty} \phi^{(n)}\!
\left[\; \sum_{n=0}^{\infty}
J^{(n)}[\phi] \; \right]
\\
&=&
\phi^{(0)}\!\left[ J^{(0)}[\phi] + J^{(1)}[\phi] +\cdots\right]
+
\phi^{(1)}\!\left[ J^{(0)}[\phi] + J^{(1)}[\phi] +\cdots\right]
+\cdots
\end{eqnarray}
or
\begin{eqnarray}
\phi&=&
\phi^{(0)}\!\left[ J^{(0)}[\phi] \right]
+
\phi^{(0)\prime}\left[ J^{(0)}[\phi]\right]J^{(1)}[\phi]
+\cdots+
\phi^{(1)}\!\left[ J^{(0)}[\phi] \right]
+\cdots
\label{im}
\end{eqnarray}
where
$\phi^{(0)\prime}[J]=\frac{\delta \phi^{(0)}[J]}{\delta J}$.
The inversion is made {\em by regarding $\phi$
as independent of} $\lambda$, namely,
as of order $\lambda^0=1$.
Then an explicit form for $J^{(n)}[\phi]$
is known successively up to the desired $n$
by writing down the $n$-th order of (\ref{im});
$\phi= \phi^{(0)}[ J^{(0)}[\phi] ],
J^{(1)}=-\phi^{(1)}[J^{(0)}[\phi]]/
\phi^{(0)\prime}[ J^{(0)}[\phi]],\ldots.$
Regarding $\phi$ as independent of $\lambda$ just corresponds to
making the Legendre transformation from $J$ to $\phi$
(see Appendix A).
The extension of the above formula to the
case of local variables $J(x)$ and $\phi(x)$ can be
done merely by recovering the $x$-dependence and appropriate
space-time integrals.
We will see that the series expansion (\ref{im}) in the graphical
form is directly given by (\ref{PHI2}) below.
An explicit form of $J^{(0)}[\phi]$ may not be obtainable in
the cases studied in this paper
because $J^{(0)}[\phi]$
is defined by the inverse of a known functional
$\phi^{(0)}[J]$
or
$J^{(0)}[\phi]=\phi^{(0)\; -1}[\phi]$.
However, examples in which $J^{(0)}[\phi]$
is explicitly obtained are dealt with in Appendix C.
But this is not necessarily an obstacle or rather may be a merit
in actual calculation in some cases.
An explicit instance in this respect has been provided for the
case of the itinerant electron model (see \cite{IEM}).
In other cases it is more convenient to change the dynamical
variable; $\phi \rightarrow J^{(0)}[\phi]$ as in \cite{SUP}.

In Sec.~2 the case of the $\varphi^4$ theory is discussed in
detail as the simplest example and also as a prototype for the
subsequent two models.
First we try to deduce the rule and
arrive at the propositions to be proved later.
Explicit rules are given in the form of
Proposition A$2$) with A$1'$) or Proposition A$3'$) below.
In the second subsection we rigorously prove these propositions
in two ways;
by use of a topological relation and  by the sum-up rule\cite{KOBS}.
In Sec.~3 the case of the itinerant model is studied
as an example of the free energy of the condensed matter physics.
More model specific study of the case has been carried
out \cite{IEM} to give a systematic improvement of the Stoner
theory and to obtain the results similar
to the SCR theory by Moriya and Kawabata \cite{MK}.
Last example of QED is given in Sec.~4 which may be a one-step
toward a gauge-invariant study of the gauge field theory.
Appendix A explains the reason why $\phi$ is to be considered
as independent of $\lambda$ in the process of inversion
in a way different from the one given in the literature.
In Appendix B the Feynman rules which are necessary
for our discussion are given in detail
because the symmetry factors play an important role
in the {\em deduction} of the rule (although they are less
important in the {\em proof} of the rule).
Appendix C reproduces the known rules of the effective actions for an
elementary field and non-local 2-body composite operators
by the inversion method.
In these cases $J^{(0)}[\phi]$ can be explicitly
given as stated before.
In Appendix D we review the path-integral technique for the fermion
coherent state used in Sec.~3.
\setcounter{equation}{0}
\section{Case of $\varphi^4$ theory}
\label{s2}
As the simplest example we consider the effective action
for the expectation value of a {\em local composite
operator} $\varphi(x)^{2}$ in the $\varphi^{4}$
theory ---
we take
$\Gamma[\phi]$ with the local variable $\phi(x) \propto
\langle \varphi(x)^{2} \rangle$.

Let us introduce the generating functional $W[J]$ in the
path-integral representation
as follows;
\begin{equation}
e^{iW[J]}=\int {\cal D} \varphi e^{iS[\varphi,J]},
\label{1}
\end{equation}
\begin{equation}
S[\varphi,J] =
- \frac{1}{2} \int d^{4} x d^{4} y
\varphi(x) G^{-1}(x,y) \varphi(y)
- \frac{\lambda}{4!} \int d^{4} x \varphi(x)^{4}
+ \frac{1}{2} \int d^{4} x
J(x) \varphi(x)^{2},
\label{2}
\end{equation}
\begin{equation}
G^{-1}(x,y) =
(\Box+m^{2})\delta^{4}(x-y),
\end{equation}
where $\int {\cal D}\varphi$
denotes the functional path-integration by the field $\varphi$.
Note here that an $x$-dependent local external source $J(x)$ is
coupled to the local composite field operator
$\varphi(x)^{2}$.
Hereafter we frequently use the notation in which the space-time
indices and their integrations are omitted if it causes no
ambiguity.
For example $S[\varphi,J]$ in (\ref{2}) is denoted as
\[
-\frac{1}{2}\varphi G^{-1} \varphi
-\frac{\lambda}{4!}\varphi^4
+\frac{1}{2}J\phi
\]
in this symbolic notation.

It is straightforward to get the graphical rule for $W[J]$.
We note here that different rules are obtained depending
on how much part of $J$ is absorbed in the propagator.
In this paper both the following two diagrammatic rules
(\ref{W1}) and (\ref{W2}) are used;
\begin{equation}
\label{W1}
W[J]-W_0=-\frac{1}{2i}{\rm Tr} \ln G_J^{-1}
+\frac{1}{i} \left\langle
e^{-\frac{i\lambda}{4!}\varphi^4 }
\right\rangle_{G_J},
\end{equation}
that is, the sum of
all the connected vacuum graphs built with the 4-point
vertex $-\lambda$ and the propagator $G_J$,
and
\begin{equation}
\label{W2}
W[J]-W_0=-\frac{1}{2i}{\rm Tr} \ln [G^{(0)}]^{-1}
+\frac{1}{i} \left\langle
e^{-\frac{i\lambda}{4!}\varphi^4
+\frac{i}{2}(J^{(1)}+ J^{(2)}+\cdots )
\varphi^2
}
\right\rangle_{G^{(0)}},
\end{equation}
that is, the sum of
all the connected vacuum graphs constructed out of the 4-point
vertex $-\lambda$, the 2-point vertex $J^{(i)}$ with
$i \geq 1$, and
the propagator $G^{(0)}$.
Here the propagators are defined as (with obvious symbolic notation)
\begin{equation}
G_J^{-1}=\Box+m^2-J
\quad
\mbox{and}
\quad
[G^{(0)}]^{-1}=\Box+m^2-J^{(0)}.
\end{equation}
$W_0$ is the trivial $J$-independent part of $W$
and Tr represents the functional trace.
The first term on the right-hand
side of (\ref{W1}) or (\ref{W2}) (${\rm Tr}\ln$ term), is
usually denoted by a circle in graphical representation
and, in this paper,
is called {\em a trivial skeleton} (the definition of
the skeleton itself is given below).
Furthermore the notation of the form
$\langle O[\varphi] \rangle_A$ means the summation of all
the possible connected Wick contraction of the operators
contained in $O[\varphi]$ by using $A$ as propagators, that is,
\begin{equation}
\langle O[\varphi] \rangle_A =
\left.
\frac{\int {\cal D} \varphi e^{iS_0}O[\varphi]}{
\int {\cal D} \varphi e^{iS_0}}\right|_{\rm conn.}
\quad
\mbox{with}
\quad
S_0=-\frac{1}{2}\varphi A^{-1} \varphi.
\label{O}
\end{equation}
Throughout this paper we employ this notation frequently
from which {\em the weights of graphs} are explicitly obtained.
Remember that the original notation
$\langle \varphi(x)^2 \rangle$ and
$\langle \varphi(x)^2 \rangle^J$
implies, however, the full order expectation value of course.
The rule (\ref{W2}) contains 2-point
vertices of $J^{(i)}\varphi^2$ ($i \geq 1$)
because the absorption of $J$ into the propagator is not
complete.

Now the expectation value of the local composite field
will be called $\phi(x)$, specifically :
\begin{equation}
\phi(x)=\frac{\delta W}{\delta J(x)}
\equiv \frac{1}{2}\langle \varphi(x)^{2} \rangle^J.
\label{6}
\end{equation}
With the notation (\ref{O}) the graphical rules
corresponding to (\ref{W1}) and (\ref{W2})
are summarized as
\begin{equation}
\label{PHI1}
\phi=
\left\langle
\frac{1}{2}\varphi^2
e^{-\frac{i\lambda}{4!}\varphi^4 }
\right\rangle_{G_J},
\end{equation}
that is, the sum of all the connected graphs with one external
point (where two propagators meet)
built with the 4-point vertex $-\lambda$
and the propagator $G_J$, and
\begin{equation}
\label{PHI2}
\phi=
\left\langle
\frac{1}{2}\varphi^2
e^{-\frac{i\lambda}{4!}\varphi^4
+\frac{i}{2}(J^{(1)}+ J^{(2)}+\cdots)
\varphi^2
}
\right\rangle_{G^{(0)}},
\end{equation}
that is, the sum of all the connected graphs with one external
point (where two propagators meet)
built with the 4-point vertex $-\lambda$,
the 2-point vertex $J^{(i)}$ ($i\geq 1$),
and the propagator $G^{(0)}$.

To rewrite the theory in terms of this dynamical
variable $\phi$ instead of $J$ we introduce as usual
the effective action of $\phi$ through Legendre
transformation:
\begin{equation}
\Gamma[\phi]=W[J]- \int d^{4} x J(x) \phi(x)
\equiv W[J]-J\phi
\label{7}
\end{equation}
with an identity
\begin{equation}
-J(x)=\frac{\delta \Gamma[\phi]}{\delta \phi(x)}.
\label{8}
\end{equation}

It is convenient to introduce $\Gamma^{(n)}$, which is
the $n$-th order in $\lambda$, or
\begin{equation}
\Gamma=\sum_{n=0}^{\infty}
\Gamma^{(n)}.
\end{equation}
Then we see in Sec.~2.1.2 that
$\Gamma^{(0)}$ and $\Gamma^{(1)}$ are explicitly given by
, suppressing the space-time integration;
\begin{equation}
\Gamma^{(0)}
=
-J^{(0)}[\phi]\phi
-\frac{1}{2i}{\rm Tr}\ln[G^{(0)}]^{-1},
\label{GAM0}
\end{equation}
\begin{equation}
\Gamma^{(1)}
=
-\frac{1}{2}\lambda\phi^2.
\label{GAM1}
\end{equation}
In this case of the $\varphi^4$ theory,
$J^{(0)}[\phi]$ is defined through
\begin{equation}
\label{0th}
\phi(x)=\frac{1}{2i} G^{(0)}(x,x)
=\frac{1}{2i}\left(\frac{1}{\Box+m^2-J^{(0)}[\phi]}\right)_{xx},
\end{equation}
which is to be proved in Sec.~2.1.1.
We emphasize here that although
the right-hand side is denoted by a single
graph of (\ref{16'}) below, $\phi$ on the left-hand
side is a full-order quantity, suggesting that $J^{(0)}[\phi]$
has a full-order  information.
The central part of our study is that for the remaining part
of $\Gamma$, which is called $\Delta \Gamma$,
\begin{equation}
\Delta \Gamma
= \sum_{i=2}^{\infty}\Gamma^{(i)}[\phi].
\end{equation}

\subsection{Perturbative derivation of the graphical
rule for $\Gamma [ \phi]$ through inversion method}
An explicit calculation up to the fourth order of $\lambda$
is sketched and based on the result the general rule
for full order is deduced.
Full justification is given in Sec.~2.2.
In Sec.~2.1.1 the rule for $J^{(n)}$ is inferred by use of
the inversion method.
We see that $J^{(n)}$ is successively given as a
functional of $J^{(0)}[\phi]$.
Then in Sec.~2.1.2 we obtain $\Gamma^{(n)}$ based on
$J^{(n)}$ vertex in two ways;
by integrating the diagrams of $J^{(n)}$ or by
starting from a closed formula for $\Delta \Gamma$.
Since $J^{(n)}$ has already been given as a functional
of $J^{(0)}[\phi]$ in Sec.~2.1.1,
the effective action $\Delta \Gamma$
is obtained as a functional of $J^{(0)}[\phi]$.
Explicit rules for $\Delta J$ and $\Delta \Gamma$ are
given in Sec.~2.1.3 in which their dependence on
$J^{(0)}[\phi]$ is transparent.
For this purpose an artificial
bosonic field $\sigma$ whose propagator is a
functional of $J^{(0)}[\phi]$ is introduced.
\subsubsection{Rule for $J^{(n)}$}
The original series of $\phi$ is first calculated as
\begin{equation}
\phi =
\phi^{(0)}+
\phi^{(1)}+
\phi^{(2)}+
\phi^{(3)}+
\phi^{(4)}+\cdots
\label{9}
\end{equation}
by (\ref{PHI1}) regarding $J$ as order unity
with graphical representation as follows;\\
\begin{equation}
\phi^{(0)}[J]=
\setlength{\unitlength}{1mm}
\begin{picture}(12,10)
\thicklines
\put(6,1){\circle{8}}
\put(1.75,1){\circle*{1}}
\end{picture} ,
\label{10}
\end{equation}
\begin{equation}
\phi^{(1)}[J]=
\setlength{\unitlength}{1mm}
\begin{picture}(20.5,10)
\thicklines
\put(6,1){\circle{8}}
\put(14.5,1){\circle{8}}
\put(1.75,1){\circle*{1}}
\end{picture},
\label{11}
\end{equation}
\begin{equation}
\phi^{(2)}[J]=
\setlength{\unitlength}{1mm}
\begin{picture}(28,10)
\thicklines
\put(6,1){\circle{8}}
\put(14.5,1){\circle{8}}
\put(23,1){\circle{8}}
\put(1.75,1){\circle*{1}}
\end{picture}
+
\setlength{\unitlength}{1mm}
\begin{picture}(28,10)
\thicklines
\put(6,1){\circle{8}}
\put(14.5,1){\circle{8}}
\put(23,1){\circle{8}}
\put(14.5,5.25){\circle*{1}}
\end{picture}
+
\setlength{\unitlength}{1mm}
\begin{picture}(16.5,10)
\thicklines
\put(6,1){\circle{8}}
\put(10,1){\circle{8}}
\put(1.75,1){\circle*{1}}
\end{picture},
\label{12}
\end{equation}
\begin{eqnarray}
\lefteqn{
\phi^{(3)}[J]=
\setlength{\unitlength}{1pt}
\begin{picture}(110,24)(-7,9)
\thicklines
\put(12,12){\circle{24}}
\put(0,12){\circle*{3}}
\put(36,12){\circle{24}}
\put(60,12){\circle{24}}
\put(84,12){\circle{24}}
\end{picture}
+
\setlength{\unitlength}{1pt}
\begin{picture}(110,24)(-7,9)
\thicklines
\put(12,12){\circle{24}}
\put(36,24){\circle*{3}}
\put(36,12){\circle{24}}
\put(60,12){\circle{24}}
\put(84,12){\circle{24}}
\end{picture}
+
\setlength{\unitlength}{1pt}
\begin{picture}(90,48)(-7,9)
\thicklines
\put(12,12){\circle{24}}
\put(0,12){\circle*{3}}
\put(36,12){\circle{24}}
\put(60,12){\circle{24}}
\put(36,-12){\circle{24}}
\end{picture}
+
\setlength{\unitlength}{1pt}
\begin{picture}(90,48)(-7,9)
\thicklines
\put(12,12){\circle{24}}
\put(36,24){\circle*{3}}
\put(36,12){\circle{24}}
\put(60,12){\circle{24}}
\put(36,-12){\circle{24}}
\end{picture}
} \nonumber \\
&&+
\setlength{\unitlength}{1pt}
\begin{picture}(80,58)(-7,9)
\thicklines
\put(12,12){\circle{24}}
\put(0,12){\circle*{3}}
\put(28,12){\circle{24}}
\put(52,12){\circle{24}}
\end{picture}
+
\setlength{\unitlength}{1pt}
\begin{picture}(80,34)(-7,9)
\thicklines
\put(12,12){\circle{24}}
\put(28,24){\circle*{3}}
\put(28,12){\circle{24}}
\put(52,12){\circle{24}}
\end{picture}
+
\setlength{\unitlength}{1pt}
\begin{picture}(80,34)(-7,9)
\thicklines
\put(12,12){\circle{24}}
\put(28,12){\circle{24}}
\put(52,12){\circle{24}}
\put(64,12){\circle*{3}}
\end{picture}
+
\setlength{\unitlength}{1pt}
\begin{picture}(40,24)(-18,-3)
\thicklines
\put(0,0){\circle{24}}
\put(-12,0){\circle*{3}}
\put(12,0){\line(-5,3){18.174}}
\put(12,0){\line(-5,-3){18.174}}
\put(-6.174,10.290){\line(0,-1){20.58}}
\end{picture} ,
\label{13}
\end{eqnarray}
\begin{equation}
\phi^{(4)}[J]=
\setlength{\unitlength}{1pt}
\begin{picture}(100,24)(-10,5)
\thicklines
\put(8,8){\circle{16}}
\put(0,8){\circle*{3}}
\put(24,8){\circle{16}}
\put(40,8){\circle{16}}
\put(56,8){\circle{16}}
\put(72,8){\circle{16}}
\end{picture}
+
\mbox{(30 diagrams).}
\end{equation}
Here the black dot $\bullet$ corresponds to
an external point  where two propagators meet and to
the insertion of the operator $\varphi(x)^2$
which is effected by
the derivative with respect to $J(x)$.
Note here the relation
$\frac{\partial G(y,z)}{\partial J(x)}
=G(y,x)G(x,z).$
The propagator
$G_J(x,y)$ and the factor $-\lambda$ are associated
with a line and a 4-point vertex respectively.
(No factor is associated with a black dot.
For detailed rule including the symmetry factor, see Appendix B.)

We mention here that the diagrams of $\phi^{(n)}$
is obtained by attaching a black dot, in all
possible ways, to one of the lines in the
graphs of the $n$-th order of $W$. For example, the 31 diagrams
of $\phi^{(4)}$ is obtained through the fourth order of $W$;
\begin{eqnarray}
\lefteqn{
W^{(4)}[J]=
\setlength{\unitlength}{1pt}
\begin{picture}(100,24)(-10,5)
\thicklines
\put(8,8){\circle{16}}
\put(24,8){\circle{16}}
\put(40,8){\circle{16}}
\put(56,8){\circle{16}}
\put(72,8){\circle{16}}
\end{picture}
+
\setlength{\unitlength}{1pt}
\begin{picture}(84,24)(-10,5)
\thicklines
\put(8,8){\circle{16}}
\put(24,8){\circle{16}}
\put(40,8){\circle{16}}
\put(56,8){\circle{16}}
\put(40,-8){\circle{16}}
\end{picture}
+
\setlength{\unitlength}{1pt}
\begin{picture}(65,24)(-10,5)
\thicklines
\put(8,8){\circle{16}}
\put(24,8){\circle{16}}
\put(40,8){\circle{16}}
\put(24,-8){\circle{16}}
\put(24,24){\circle{16}}
\end{picture}
+
\setlength{\unitlength}{1pt}
\begin{picture}(77,24)(-10,5)
\thicklines
\put(8,8){\circle{16}}
\put(24,8){\circle{16}}
\put(34,8){\circle{16}}
\put(50,8){\circle{16}}
\end{picture}
} \nonumber \\
&&+
\setlength{\unitlength}{1pt}
\begin{picture}(60,45)(-10,5)
\thicklines
\put(8,8){\circle{16}}
\put(18,8){\circle{16}}
\put(34,8){\circle{16}}
\put(18,-8){\circle{16}}
\end{picture}
+
\setlength{\unitlength}{1pt}
\begin{picture}(76,45)(-10,5)
\thicklines
\put(8,8){\circle{16}}
\put(24,8){\circle{16}}
\put(40,8){\circle{16}}
\put(50,8){\circle{16}}
\end{picture}
+
\setlength{\unitlength}{1pt}
\begin{picture}(49,24)(0,-3)
\thicklines
\put(17,0){\circle{24}}
\put(29,0){\circle{24}}
\put(35,0){\circle{12}}
\end{picture}
+
\setlength{\unitlength}{1pt}
\begin{picture}(51,34)(5,-3)
\thicklines
\put(17,0){\circle{16}}
\put(29,0){\circle{24}}
\put(41,0){\circle{16}}
\end{picture}
+
\setlength{\unitlength}{1pt}
\begin{picture}(56,24)(-36,-3)
\thicklines
\put(-20,0){\circle{16}}
\put(0,0){\circle{24}}
\put(12,0){\line(-5,3){18.174}}
\put(12,0){\line(-5,-3){18.174}}
\put(-6.174,10.290){\line(0,-1){20.58}}
\end{picture}
+
\setlength{\unitlength}{1pt}
\begin{picture}(33,34)(-18,-3)
\thicklines
\put(0,0){\circle{24}}
\put(8.485,8.485){\line(0,-1){16.97}}
\put(8.485,8.485){\line(-1,0){16.97}}
\put(-8.485,-8.485){\line(0,1){16.97}}
\put(-8.485,-8.485){\line(1,0){16.97}}
\end{picture} .
\label{14}
\end{eqnarray}

Since the above diagrams of $\phi^{(i)}$
are all functional of $J(x)$,
which is contained in the propagator $G_J(x,y)$, we get $\phi(x)$
as a functional of $J$; $\phi=\phi[J]$.
Assume that the relation $\phi=\phi[J]$ is {\em inverted}
to give the relation $J=J[\phi]$ and this inversion is
done perturbatively as in (\ref{1.2}) regarding $\phi$ as an quantity
independent of $\lambda$ or the order $\lambda^{0}=1$.
Then as in
Introduction
we get the following formulae of the
inversion method;
\begin{equation}
\phi=\phi^{(0)}[J^{(0)}],
\label{16} \\
\end{equation}
\begin{equation}
\phi^{(0)\prime}J^{(1)}+\phi^{(1)}=0,
\label{17} \\
\end{equation}
\begin{equation}
\phi^{(0)\prime}J^{(2)}+
\frac{1}{2}\phi^{(0)\prime\prime}
\left(J^{(1)}\right)^{2}+
\phi^{(1)\prime}J^{(1)}+\phi^{(2)}=0,
\label{18} \\
\end{equation}
\begin{eqnarray}
\lefteqn{
\phi^{(0)\prime}J^{(3)}+\phi^{(0)\prime\prime}
J^{(1)}J^{(2)}+
\frac{1}{3!}\phi^{(0)\prime\prime\prime}
\left(J^{(1)}\right)^{3}
}
\nonumber \\
&+&
\phi^{(1)\prime}J^{(2)}+
\frac{1}{2}\phi^{(1)\prime\prime}\left(J^{(1)}\right)^{2}+
\phi^{(2)\prime}J^{(1)}+\phi^{(3)}=0,
\label{19}
\end{eqnarray}
\begin{eqnarray}
\lefteqn{
\phi^{(0)\prime}J^{(4)}+
\frac{1}{2}\phi^{(0)\prime\prime}\left(2J^{(1)}J^{(3)}
+
\left(J^{(2)}\right)^2 \right)+
\frac{1}{2}\phi^{(0)\prime\prime\prime}
\left(J^{(1)}\right)^2 J^{(2)}+
\frac{1}{4!}\phi^{(0)\prime\prime\prime\prime}
\left(J^{(1)}\right)^{4}
}
\nonumber \\
&+&
\phi^{(1)\prime}J^{(3)}+
\phi^{(1)\prime\prime}J^{(1)}J^{(2)}+
\frac{1}{3!}\phi^{(1)\prime\prime\prime}
\left(J^{(1)}\right)^{3}
\nonumber \\
&+&
\phi^{(2)\prime}J^{(2)}+
\frac{1}{2}\phi^{(2)\prime\prime}
\left(J^{(1)}\right)^{2}+
\phi^{(3)\prime}J^{(1)}+\phi^{(4)}=0.
\label{20}
\end{eqnarray}
Here we have employed a concise notation. If we explicitly write
(\ref{18}), for example, it has the form;
\begin{eqnarray}
\lefteqn{
\int d^{4} x
\frac{\delta \phi^{(0)}[J^{(0)}]}{\delta J^{(0)}(x)}
J^{(2)}(x)+
\frac{1}{2} \int d^{4} x d^{4} y
\frac{\delta \phi^{(0)}[J^{(0)}]}{
\delta J^{(0)}(x)\delta J^{(0)}(y)}
J^{(1)}(x)J^{(1)}(y)
}
\nonumber \\
 &+&
\int d^{4} x
\frac{\delta \phi^{(1)}[J^{(0)}]}{\delta J^{(0)}(x)}
J^{(1)}(x)+\phi^{(2)}[J^{(0)}] =0.
\label{18'}
\end{eqnarray}
We emphasize here that all $\phi^{(i)}$
($i=0,1,2,\ldots$) and their derivatives in (\ref{16}) to (\ref{20})
are evaluated at $J=J^{(0)}[\phi]$ defined implicitly by (\ref{16}).
So equations (\ref{17}) to (\ref{20})
successively give the functional dependence of
$J^{(1)}$ to
$J^{(4)}$
on $\phi$
{\em through $J^{(0)}[\phi]$}.

Let us discuss the graphical expressions
of (\ref{16}) to (\ref{20}).
Note here that the propagator
in the following graphs is
$ G^{(0)} = \frac{1}{\Box+m^2-J^{(0)}[\phi]}$
instead of $G_J$.
Then, from (\ref{10}), eq.~(\ref{16}) is expressed as
\begin{equation}
\phi=
\setlength{\unitlength}{1mm}
\begin{picture}(13,10)
\thicklines
\put(6,1){\circle{8}}
\put(1.75,1){\circle*{1}}
\end{picture}.
\label{16'}
\end{equation}
Here and hereafter the black dot
represents  derivative not by
$J$ but by $J^{(0)}$.
Notice also that the meanings of the graphs
on the right-hand side of (\ref{10}) and (\ref{16'})
are different because the line or the propagator
in them is not the same:
$G_J$ for (\ref{10}) and $G^{(0)}$ for (\ref{16'}).
Thus (\ref{16'}) reduces to (\ref{0th}).
It is stressed here that $J^{(0)}[\phi]$ is defined through
(\ref{0th}) or (\ref{16'})
although its dependence on $\phi$ is only implicit.
By using (\ref{10}) and (\ref{11}), eq.~(\ref{17})
is also expressed as follows.
\begin{equation}
\setlength{\unitlength}{1mm}
\begin{picture}(16,10)
\thicklines
\put(4,1){\circle{8}}
\put(-0.25,1){\circle*{1}}
\put(8.25,1){\circle*{1}}
\put(12,1){\makebox(0,0){$J^{(1)}$}}
\end{picture}
+
\setlength{\unitlength}{1mm}
\begin{picture}(20,10)
\thicklines
\put(6,1){\circle{8}}
\put(14.5,1){\circle{8}}
\put(1.75,1){\circle*{1}}
\end{picture}
=0.
\label{17'}
\end{equation}
Here we have used the relation
\begin{equation}
\phi^{(0)\prime}=
\setlength{\unitlength}{1mm}
\begin{picture}(12,10)(-2,0)
\thicklines
\put(4,1){\circle{8}}
\put(-0.25,1){\circle*{1}}
\put(8.25,1){\circle*{1}}
\end{picture}.
\end{equation}
Noting that a 4-point vertex makes a
contribution $-\lambda$ so that
\begin{equation}
\setlength{\unitlength}{1mm}
\begin{picture}(19,10)
\thicklines
\put(6,1){\circle{8}}
\put(14.5,1){\circle{8}}
\put(1.75,1){\circle*{1}}
\end{picture}
=
\setlength{\unitlength}{1mm}
\begin{picture}(12,10)(-2,0)
\thicklines
\put(4,1){\circle{8}}
\put(-0.25,1){\circle*{1}}
\put(8.25,1){\circle*{1}}
\end{picture}
(-\lambda)
\setlength{\unitlength}{1mm}
\begin{picture}(12,10)(-2,0)
\thicklines
\put(4,1){\circle{8}}
\put(-0.25,1){\circle*{1}}
\end{picture},
\end{equation}
we get from (\ref{17'})
\begin{equation}
J^{(1)}=\lambda
\setlength{\unitlength}{1mm}
\begin{picture}(16,10)(-2,0)
\thicklines
\put(4,1){\circle{8}}
\put(-0.25,1){\circle*{1}}
\end{picture}
\label{21}
\end{equation}
or
\begin{equation}
J^{(1)}= \lambda \phi =
\lambda \frac{1}{2i} {\rm Tr}
\frac{1}{\Box+m^2-J^{(0)}[\phi]}.
\label{22}
\end{equation}
Thus $J^{(1)}$ is given by $J^{(0)}[\phi]$.
Consider next the graphical expression of (\ref{18})
obtained through (\ref{10}) to (\ref{12}).
\begin{eqnarray}
\lefteqn{
\setlength{\unitlength}{1mm}
\begin{picture}(19,10)(-2,0)
\thicklines
\put(4,1){\circle{8}}
\put(-0.25,1){\circle*{1}}
\put(8.25,1){\circle*{1}}
\put(12,1){\makebox(0,0){$J^{(2)}$}}
\end{picture}
+
\setlength{\unitlength}{1pt}
\begin{picture}(50,24)(-18,-3)
\thicklines
\put(-12,0){\circle*{3}}
\put(0,0){\circle{24}}
\put(8.485,8.485){\circle*{3}}
\put(8.485,-8.485){\circle*{3}}
\put(11,10){\makebox(0,0)[l]{$J^{(1)}$}}
\put(8.485,-8.485){\makebox(0,0)[l]{$J^{(1)}$}}
\end{picture}
+
\setlength{\unitlength}{1pt}
\begin{picture}(75,24)(0,-3)
\thicklines
\put(17,0){\circle{24}}
\put(5,0){\circle*{3}}
\put(41,0){\circle{24}}
\put(53,0){\circle*{3}}
\put(53,0){\makebox(0,0)[l]{$J^{(1)}$}}
\end{picture}
+
\setlength{\unitlength}{1pt}
\begin{picture}(65,24)(0,-3)
\thicklines
\put(17,0){\circle{24}}
\put(5,0){\circle*{3}}
\put(17,12){\circle*{3}}
\put(41,0){\circle{24}}
\put(17,15){\makebox(0,0)[b]{$J^{(1)}$}}
\end{picture}
} \nonumber \\
&&+
\setlength{\unitlength}{1pt}
\begin{picture}(81,24)(0,-3)
\thicklines
\put(17,0){\circle{24}}
\put(5,0){\circle*{3}}
\put(41,0){\circle{24}}
\put(65,0){\circle{24}}
\end{picture}
+
\setlength{\unitlength}{1pt}
\begin{picture}(60,24)(0,-3)
\thicklines
\put(17,0){\circle{24}}
\put(5,0){\circle*{3}}
\put(41,0){\circle{24}}
\put(17,-24){\circle{24}}
\end{picture}
+
\setlength{\unitlength}{1pt}
\begin{picture}(50,34)(0,-3)
\thicklines
\put(17,0){\circle{24}}
\put(5,0){\circle*{3}}
\put(30,0){\circle{24}}
\end{picture}
= 0.
\label{23}
\end{eqnarray}
\\
We see that the second, fourth, and sixth graphs on the left-hand side
are summed up to zero
after replacing $J^{(1)}$ by the right-hand side of  (\ref{21})
by explicitly taking symmetry factors into account of course
 -- see Appendix B.
A similar cancellation of the
third and fifth graphs on the left-hand side of (\ref{23})
occurs ending up with
\begin{equation}
\label{J2}
-
\setlength{\unitlength}{1mm}
\begin{picture}(19,10)(-2,0)
\thicklines
\put(4,1){\circle{8}}
\put(-0.25,1){\circle*{1}}
\put(8.25,1){\circle*{1}}
\put(12,1){\makebox(0,0){$J^{(2)}$}}
\end{picture}
=
\setlength{\unitlength}{1mm}
\begin{picture}(15,10)
\thicklines
\put(5,1){\circle{8}}
\put(0.75,1){\circle*{1}}
\put(9,1){\circle{8}}
\end{picture} .
\end{equation}
The graphs of (\ref{19}) and (\ref{20}) are also
obtained through (\ref{10}) to (\ref{14}).
These expressions originally consist of
many terms but due to
similar cancellation mechanism, they reduce to
\begin{equation}
-
\setlength{\unitlength}{1mm}
\begin{picture}(19,10)(-2,0)
\thicklines
\put(4,1){\circle{8}}
\put(-0.25,1){\circle*{1}}
\put(8.25,1){\circle*{1}}
\put(12,1){\makebox(0,0){$J^{(3)}$}}
\end{picture}
\label{J3}
=
\setlength{\unitlength}{1pt}
\begin{picture}(36,24)(-18,-3)
\thicklines
\put(0,0){\circle{24}}
\put(-12,0){\circle*{3}}
\put(12,0){\line(-5,3){18.174}}
\put(12,0){\line(-5,-3){18.174}}
\put(-6.174,10.290){\line(0,-1){20.58}}
\end{picture} ,
\end{equation}
\begin{eqnarray}
\lefteqn{
-
\setlength{\unitlength}{1mm}
\begin{picture}(19,10)(-2,0)
\thicklines
\put(4,1){\circle{8}}
\put(-0.25,1){\circle*{1}}
\put(8.25,1){\circle*{1}}
\put(12,1){\makebox(0,0){$J^{(4)}$}}
\end{picture}
=
\setlength{\unitlength}{1pt}
\begin{picture}(50,24)(-15,-3)
\thicklines
\put(-12,0){\circle*{3}}
\put(0,0){\circle{24}}
\put(8.485,8.485){\circle*{3}}
\put(8.485,-8.485){\circle*{3}}
\put(11,10){\makebox(0,0)[l]{$J^{(2)}$}}
\put(8.485,-8.485){\makebox(0,0)[l]{$J^{(2)}$}}
\end{picture}
+
\setlength{\unitlength}{1pt}
\begin{picture}(65,24)(-12,-3)
\thicklines
\put(17,0){\circle{24}}
\put(8.515,8.485){\circle*{3}}
\put(29,0){\circle{24}}
\put(5,0){\circle*{3}}
\put(10,14){\makebox(0,0)[r]{$J^{(2)}$}}
\end{picture}
} \nonumber \\
&&+
\setlength{\unitlength}{1pt}
\begin{picture}(65,24)(0,-3)
\thicklines
\put(17,0){\circle{24}}
\put(5,0){\circle*{3}}
\put(29,0){\circle{24}}
\put(41,0){\circle*{3}}
\put(42,0){\makebox(0,0)[l]{$J^{(2)}$}}
\end{picture}
+
\setlength{\unitlength}{1pt}
\begin{picture}(49,24)(0,-3)
\thicklines
\put(5,0){\circle*{3}}
\put(17,0){\circle{24}}
\put(29,0){\circle{24}}
\put(35,0){\circle{12}}
\end{picture}
+
\setlength{\unitlength}{1pt}
\begin{picture}(49,24)(0,-3)
\thicklines
\put(29,12){\circle*{3}}
\put(17,0){\circle{24}}
\put(29,0){\circle{24}}
\put(35,0){\circle{12}}
\end{picture}
+
\setlength{\unitlength}{1pt}
\begin{picture}(49,24)(0,-3)
\thicklines
\put(35,6){\circle*{3}}
\put(17,0){\circle{24}}
\put(29,0){\circle{24}}
\put(35,0){\circle{12}}
\end{picture}
+
\setlength{\unitlength}{1pt}
\begin{picture}(51,24)(5,-3)
\thicklines
\put(9,0){\circle*{3}}
\put(17,0){\circle{16}}
\put(29,0){\circle{24}}
\put(41,0){\circle{16}}
\end{picture}
+
\setlength{\unitlength}{1pt}
\begin{picture}(51,24)(5,-3)
\thicklines
\put(29,12){\circle*{3}}
\put(17,0){\circle{16}}
\put(29,0){\circle{24}}
\put(41,0){\circle{16}}
\end{picture}
+
\setlength{\unitlength}{1pt}
\begin{picture}(34,34)(-18,-3)
\thicklines
\put(-12,0){\circle*{3}}
\put(0,0){\circle{24}}
\put(8.485,8.485){\line(0,-1){16.97}}
\put(8.485,8.485){\line(-1,0){16.97}}
\put(-8.485,-8.485){\line(0,1){16.97}}
\put(-8.485,-8.485){\line(1,0){16.97}}
\end{picture}.
\nonumber \\
\label{J4}
\end{eqnarray}
These simple results lead us to the following proposition
to be justified later.
Before presenting the proposition it is convenient
to introduce the terms 1VI and 1VR.
The 1VI (1-vertex-irreducible) graph is a
connected graph in which removal
of any one of the 4-point vertices does not lead
to two separate graphs.
The 1VR (1-vertex-reducible) vertex is defined as
a 4-point vertex in a connected diagram deletion
of which results in a separation of the graph.
The 1VI graph can also be defined as the connected
graph without any 1VR vertex while
the 1VR graph is a graph in which at least one 1VR vertex is
present.
By definition {\em a graph which does not have any 4-point
vertex is not 1VR but 1VI} although the trivial skeleton
(${\rm Tr}\ln$ term) is not 1VR nor 1VI.
Namely all the graphs are classified into three categories;
1VI, 1VR, and the trivial skeleton.
For later convenience we introduce the skeleton. Both the
1VI graph and the trivial skeleton is called the skeleton.
In other words
the whole class of the skeleton is all
the 1VI graphs plus the trivial skeleton.
With this terminology we see that all the 1VR graph in
(\ref{23}) disappear to result in (\ref{J2})
after all the $J^{(1)}$'s are replaced
by the right-hand side of (\ref{21}).
Thus we can deduce the following proposition.
\begin{description}
\item[Proposition A$\bf 1$)]
After replacing $J^{(1)}$ by its graphical expression
of the right-hand side of
(\ref{21}), all the 1VR graphs
originally appearing in the inversion formula
of the $n$-th order with $n\geq 2$ ((\ref{18})
to (\ref{20}) and higher relations) cancel out.
In other words, only the 1VI
graphs with correct weight remain in the inversion
formulae.
\end{description}
Note here that 1VI can not be replaced by
2PI (2-particle-irreducible) as in the case of
the effective action for the
{\em non-local} operator $\varphi(x)\varphi(y)$\cite{DM,VKP,CJT}.
This is clear from the second and third (1VI) graphs from the
last on the left-hand side of (\ref{J4}), which are 2PR
(2-particle-reducible).

We also note a very convenient way to express the {\em original}
graphs of the inversion formulae
(\ref{16}) to (\ref{20}) and higher order relations
such as (\ref{16'}), (\ref{17'}), (\ref{23})
in which graphs $J^{(1)}$'s still remain
(without replacing them by the right-hand side of (\ref{21})).
Let us turn our attention to (\ref{PHI2})
where graphs are built with
propagators $G^{(0)}$ and 4-point vertices $-\lambda$
and {\em pseudo-vertices} of order $\lambda^i$ with $i\geq 1$,
which is denoted as
\setlength{\unitlength}{1 mm}
\begin{picture}(25,10)
\thicklines
\put(0,0){\line(1,0){20}}
\put(10,0){\circle*{1}}
\put(10,1){\makebox(0,0)[b]{$J^{(i)}$}}
\put(23,0){\makebox(0,0){.}}
\end{picture}
We have called  the 2-point vertex originating from
$J^{(n)}\varphi^2$ a {\em pseudo-vertex}
since it has nothing to do with the definition of 1VI.
The term 1VI is defined as 1-vertex-irreducible with respect to
4-point vertex.
Then the graphs of the inversion formula are obtained as
follows.
{\em If one writes down the n-th order of (\ref{PHI2}) considering
$\phi$ and $G^{(0)}$ (namely, $J^{(0)}$) as of order $\lambda^0=1$
one obtains the inversion formula of order $n$ in the
graphical form}.
For example, the $0$-th order of (\ref{PHI2}) is
\begin{equation}
\phi=\left\langle
\frac{1}{2}\varphi^2
\right\rangle_{G^{(0)}},
\label{C}
\end{equation}
which is equivalent to (\ref{16'}) and the first order is
\begin{equation}
0= \left\langle
\frac{1}{2}\varphi^2
\left(
-\frac{i\lambda}{4!}\varphi^4+\frac{i}{2}
J^{(1)}\varphi^2
\right)
\right\rangle_{G^{(0)}},
\label{D}
\end{equation}
which is (\ref{17'}).
Furthermore the second order of (\ref{PHI2}) reduces
to (\ref{23}).
Here it is convenient
to introduce {\em the self-contraction}
of the pseudo-vertex (Fig.~1) and the 4-point vertex (Fig.~2).
Since the relation
$\frac{\delta G^{(0)}}{\delta J^{(0)}}=G^{(0)}G^{(0)}$
holds the quantity
\begin{equation}
\setlength{\unitlength}{1mm}
\begin{picture}(17,10)(-2,0)
\thicklines
\put(4,1){\circle{8}}
\put(-0.25,1){\circle*{1}}
\put(8.25,1){\circle*{1}}
\put(12,1){\makebox(0,0){$J^{(n)}$}}
\end{picture}
=
\left\langle
\frac{1}{2}\varphi^2
\cdot
\frac{i}{2} J^{(n)}\varphi^2
\right\rangle_{G^{(0)}}
=
\frac{1}{2i}
\frac{1}{\Box+m^2-J^{(0)}}
\frac{1}{\Box+m^2-J^{(0)}}
J^{(n)}
=
\frac{\delta}{\delta J^{(0)}}
\setlength{\unitlength}{1mm}
\begin{picture}(19,10)(-2,0)
\thicklines
\put(4,1){\circle{8}}
\put(8.25,1){\circle*{1}}
\put(12,1){\makebox(0,0){$J^{(n)}$}}
\end{picture}
\label{B}
\end{equation}
is called (the $n$-th order of) {\em the derivative of
the self-contraction}\/ in the following.
Then as we have, from (\ref{PHI2}),
\begin{equation}
0=
\mbox{ $n$-th order of }
\left\langle
\frac{1}{2}\varphi^2
e^{-\frac{i\lambda}{4!}\varphi^4+
\frac{i}{2}(J^{(1)}+ J^{(2)}+\cdots)\varphi^2 }
\right\rangle_{G^{(0)}}
\label{tmp1}
\end{equation}
for $n\geq 1$ we get the following formula.
\begin{equation}
-
\setlength{\unitlength}{1mm}
\begin{picture}(19,10)(-2,0)
\thicklines
\put(4,1){\circle{8}}
\put(-0.25,1){\circle*{1}}
\put(8.25,1){\circle*{1}}
\put(12,1){\makebox(0,0){$J^{(n)}$}}
\end{picture}
=
\mbox{ $n$-th order of }
\left\langle
\frac{1}{2}\varphi^2
e^{-\frac{i\lambda}{4!}\varphi^4+
\frac{i}{2}(J^{(1)}+ J^{(2)}+\cdots)\varphi^2 }
\right\rangle_{G^{(0)}}^{\rm ndself.}
\label{A}
\end{equation}
where ndself.\ (no derivative of the self-contraction)
implies that the derivative of the self-contraction is moved
on the left-hand side.
Since (\ref{A}) is the original inversion formulae
of the $n$-th order
Proposition A$1$) implies that in (\ref{A})
all the contribution from the $J^{(1)}$ vertices
should be eliminated if only 1VI graphs are kept.
Thus Proposition A$1'$) follows:
\begin{description}
\item[Proposition A$\bf 1'$)]
$J^{(n)}[\phi]$ ($n\geq 2$) is successively
obtained {\em as a functional of}
$J^{(0)}[\phi]$ through
\begin{equation}
\label{J}
-
\setlength{\unitlength}{1mm}
\begin{picture}(19,10)(-2,0)
\thicklines
\put(4,1){\circle{8}}
\put(-0.25,1){\circle*{1}}
\put(8.25,1){\circle*{1}}
\put(12,1){\makebox(0,0){$J^{(n)}$}}
\end{picture}
=
\mbox{ $n$-th order of }
\left\langle
\frac{1}{2}\varphi^2
e^{-\frac{i\lambda}{4!}\varphi^4+
\frac{i}{2}(J^{(2)}+ J^{(3)}+\cdots)\varphi^2 }
\right\rangle_{G^{(0)}}^{\rm 1VI/ndself.}
\end{equation}
that is, the sum of all the connected 1VI/ndself.\/ diagrams with one
external point built with
the 4-point vertex of $-\lambda$, the 2-point pseudo-vertex
$J^{(i)}$ ($i\geq 2$) and the propagator $G^{(0)}$.
\end{description}
The restriction 1VI/ndself.\ implies that the derivative by
$J^{(0)}$ of the self-contracted diagram are excluded
and, at the same time, only the 1VI graphs should be kept.
This proposition is directly proved in the next subsection
by using the sum-up rule\cite{KOBS}.

We can {\em directly} get (\ref{J2}) to (\ref{J4})
from Proposition A$1'$) due to the 1VI/ndself.~restriction
(through the procedure similar to the one given in
(\ref{C}) or (\ref{D}) etc.).
We notice that the right-hand side of (\ref{J}) contains
$J^{(i)}$'s with $i< n$ (strictly speaking
$i\leq n-2$).
Hence we successively obtain
$J^{(i)}$ ($i\geq 2$) as  functional of $J^{(0)}$.
For example, $J^{(4)}$ is expressed by $J^{(0)}$
if one insert $J^{(2)}$ vertex given in (\ref{J2}) into (\ref{J4}).
In this way $J^{(i)}$ can be successively given as
a functional of $J^{(0)}[\phi]$
(without using $J^{(j)}$ ($j < i$, $i\neq 0$)).
\subsubsection{Rule for $\Delta \Gamma$ by use
of the pseudo-vertex $J^{(n)}$ with $n\geq 2$}
Now we turn our attention to the effective action
$\Gamma[\phi]$ itself.
We notice that $\Gamma^{(n)}[\phi]$ satisfies
\begin{equation}
\frac{\delta \Gamma^{(n)}[\phi]}{\delta \phi(x)}
=-J^{(n)}(x).
\label{27}
\end{equation}
Then for $n=0$ we get (\ref{GAM0})
because by differentiating the right-hand side of
(\ref{GAM0}) with respect to $\phi$ one obtains
$-J^{(0)}$ through (\ref{0th}).
$\Gamma^{(1)}$ is also easily obtained by
integration of (\ref{22}) so that we have (\ref{GAM1}).
To derive $\Gamma^{(n)}$ for higher $n$, it is
convenient to note the fact that
\begin{equation}
-J^{(n)}=
\frac{\delta \Gamma^{(n)}[\phi]}{\delta \phi}
=
\frac{\delta J^{(0)}[\phi]}{\delta \phi}
\frac{\delta \Gamma^{(n)}[\phi]}{\delta J^{(0)}}
\label{30}
\end{equation}
and that
\begin{equation}
\frac{\delta \phi}{\delta J^{(0)}}
=
-D^{-1}
=
\setlength{\unitlength}{1mm}
\begin{picture}(13,10)(-2,0)
\thicklines
\put(4,1){\circle{8}}
\put(-0.25,1){\circle*{1}}
\put(8.25,1){\circle*{1}}
\end{picture}.
\label{31}
\end{equation}
The quantity of the last equation is a kind of propagator
for the composite operator $\varphi(x)^2$
and is called {\em composite propagator}
($\langle \varphi(x)^2\varphi(y)^2 \rangle$).
$D$ is called the inverse composite propagator
in what follows.
Thus we get from (\ref{30})
\begin{equation}
-
\setlength{\unitlength}{1mm}
\begin{picture}(19,10)(-2,0)
\thicklines
\put(4,1){\circle{8}}
\put(-0.25,1){\circle*{1}}
\put(8.25,1){\circle*{1}}
\put(12,1){\makebox(0,0){$J^{(n)}$}}
\end{picture}
=
\frac{\delta \Gamma^{(n)}}{\delta J^{(0)}}.
\label{32}
\end{equation}
Therefore the right-hand side of (\ref{J}) is just
$\frac{\delta \Gamma^{(n)}}{\delta J^{(0)}}$.
Keeping (\ref{32}) in mind and by integrating
(\ref{J2}) to (\ref{J4}) we arrive at
\begin{equation}
\label{G2}
\Gamma^{(2)}=
\setlength{\unitlength}{1mm}
\begin{picture}(15,10)
\thicklines
\put(5,1){\circle{8}}
\put(9,1){\circle{8}}
\end{picture} ,
\end{equation}
\begin{equation}
\Gamma^{(3)}=
\setlength{\unitlength}{1pt}
\begin{picture}(40,24)(-18,-3)
\thicklines
\put(0,0){\circle{24}}
\put(12,0){\line(-5,3){18.174}}
\put(12,0){\line(-5,-3){18.174}}
\put(-6.174,10.290){\line(0,-1){20.58}}
\end{picture} ,
\label{G3}
\end{equation}
\begin{equation}
\label{G4}
\Gamma^{(4)}=
\setlength{\unitlength}{1pt}
\begin{picture}(70,24)(-35,-3)
\thicklines
\put(0,0){\circle{24}}
\put(-12,0){\circle*{3}}
\put(12,0){\circle*{3}}
\put(13,0){\makebox(0,0)[l]{$J^{(2)}$}}
\put(-14,0){\makebox(0,0)[r]{$J^{(2)}$}}
\end{picture}
+
\setlength{\unitlength}{1pt}
\begin{picture}(65,24)(0,-3)
\thicklines
\put(17,0){\circle{24}}
\put(29,0){\circle{24}}
\put(41,0){\circle*{3}}
\put(42,0){\makebox(0,0)[l]{$J^{(2)}$}}
\end{picture}
+
\setlength{\unitlength}{1pt}
\begin{picture}(49,24)(0,-3)
\thicklines
\put(17,0){\circle{24}}
\put(29,0){\circle{24}}
\put(35,0){\circle{12}}
\end{picture}
+
\setlength{\unitlength}{1pt}
\begin{picture}(51,24)(5,-3)
\thicklines
\put(17,0){\circle{16}}
\put(29,0){\circle{24}}
\put(41,0){\circle{16}}
\end{picture}
+
\setlength{\unitlength}{1pt}
\begin{picture}(40,24)(-18,-3)
\thicklines
\put(0,0){\circle{24}}
\put(8.485,8.485){\line(0,-1){16.97}}
\put(8.485,8.485){\line(-1,0){16.97}}
\put(-8.485,-8.485){\line(0,1){16.97}}
\put(-8.485,-8.485){\line(1,0){16.97}}
\end{picture} 
\end{equation}
with
\begin{equation}
\label{j2}
J^{(2)}=-
\left(
\setlength{\unitlength}{1mm}
\begin{picture}(12,4)
\thicklines
\put(6,1){\circle{8}}
\put(1.75,1){\circle*{1}}
\put(10.25,1){\circle*{1}}
\put(12.5,4){\makebox(0,0)[l]{$\scriptsize -\! 1$}}
\end{picture}
\right)
\setlength{\unitlength}{1mm}
\begin{picture}(17,10)(-2,0)
\thicklines
\put(5,1){\circle{8}}
\put(0.75,1){\circle*{1}}
\put(9,1){\circle{8}}
\end{picture}.
\end{equation}
Notice that the symmetry factors play an important role
in the integration (see Appendix B).
Note also that the
 first factor on the right-hand side of (\ref{j2}) corresponds
to {\em the amputation} of the composite propagator (\ref{31}).

In fact we can derive (\ref{G2}) to (\ref{G4}) and higher order
relations more easily.
To this end we introduce
a closed form of functional representation
of $\Delta \Gamma[\phi]$.
We first write down the following equation, which is clear from
(\ref{1}), (\ref{2}), and (\ref{7});
\begin{equation}
e^{i\Gamma[\phi]} =
\int {\cal D} \varphi
e^{i[
-\frac{1}{2}\varphi(\Box+m^2)\varphi
-\frac{1}{4!}\lambda \varphi^4
+\frac{1}{2}J \varphi^2
-J\phi
]}
\label{47}
\end{equation}
where $J$ is expressed by $\phi$.
Noting that
\begin{equation}
J=J^{(0)}[\phi]+\lambda \phi +\Delta J[\phi]
\label{48}
\end{equation}
with
\begin{equation}
\Delta J =J^{(2)}+J^{(3)}+\cdots=
-\frac{\delta \Delta \Gamma}{\delta \phi}
\label{49}
\end{equation}
and that, apart from irrelevant constant factor,
\begin{equation}
\int {\cal D} \varphi
e^{
-i\frac{1}{2}\varphi(\Box+m^2-J^{(0)})\varphi}
=
e^{
-\frac{1}{2}{\rm Tr} \ln (\Box+m^2-J^{(0)})},
\label{50}
\end{equation}
we get
\begin{equation}
e^{i\Gamma[\phi]} =
e^{i[
-J^{(0)}\phi
-\frac{1}{2i}{\rm Tr} \ln (\Box+m^2-J^{(0)})
-\frac{\lambda}{2}\phi^2
]}
\frac{
\int {\cal D} \varphi
e^{i[
-\frac{1}{2}\varphi(\Box+m^2-J^{(0)})\varphi
-\frac{1}{4!}\lambda \varphi^4
+\frac{1}{2}(J-J^{(0)}) \varphi^2
-(J-J^{(0)}-\frac{\lambda}{2}\phi)\phi
]}
}{
\int {\cal D} \varphi
e^{
-i\frac{1}{2}\varphi(\Box+m^2-J^{(0)})\varphi
}}.
\nonumber
\end{equation}
In this way a closed formula for $\Delta \Gamma$ is obtained;
\begin{equation}
e^{i\Delta \Gamma[\phi]} =
\frac{
\int {\cal D} \varphi
e^{i[
-\frac{1}{2}\varphi(\Box+m^2-J^{(0)})\varphi
+\{
-\frac{1}{4!}\lambda \varphi^4
+\frac{\lambda}{2}\phi \varphi^2
-\frac{\lambda}{2} \phi^2 \}
-\frac{\delta \Delta \Gamma}{\delta \phi}
(\frac{\varphi^2}{2}-\phi)
]}}{
\int {\cal D} \varphi
e^{
-i\frac{1}{2}\varphi(\Box+m^2-J^{(0)})\varphi
}}.
\label{51}
\end{equation}
This equation indicates that $\Delta \Gamma$ can be calculated
perturbatively by using
$G^{(0)} = (\Box+m^2-J^{(0)})^{-1}$ as propagators.
The role of {\em the additional vertices}
$\frac{\lambda}{2}\phi \varphi^2-
\frac{\lambda}{2}\phi^2$
and
$\frac{\delta \Delta \Gamma}{\delta \phi} \phi$
are merely to suppress the self-contractions.
In other words the graphs having the structure of
Fig.~1 and Fig.~2 disappear.
To see this let us take one specific
$-\frac{\lambda}{4!}\varphi^4$ vertex.
Each one of 4 $\varphi$'s of the vertex
is to be contracted with the
other $\varphi$.
There are three possible ways of such contractions.
\begin{equation}
-\frac{\lambda}{4!}\varphi^4
\Rightarrow
-\frac{\lambda}{4!}\mbox{ :}\varphi^4\mbox{: }
-\frac{\lambda}{4!}\frac{4\cdot3}{2}
\widehat{\varphi \varphi}\mbox{ :}\varphi^2\mbox{: }
-\frac{\lambda}{4!}\frac{4\cdot3}{2\cdot2}
\widehat{\varphi \varphi}
\widehat{\varphi \varphi}
\label{52}
\end{equation}
where the normal ordering
$\mbox{ :}\varphi^n\mbox{: }$
means that each one of the $n$ $\varphi$'s is to be
contracted with $\varphi$ contained in a
vertex different form the one we are taking.
Note here the contraction within a single vertex
(self-contraction) is given by
\begin{equation}
\label{53}
\widehat{\varphi \varphi}=
\frac{1}{i}\left(
\frac{1}{\Box+m^2-J^{(0)}}
\right)_{xx}
=
2\phi.
\end{equation}
In a similar manner we can write
\begin{equation}
\label{54}
\frac{\lambda}{2}\phi \varphi^2
\Rightarrow
\frac{\lambda}{2}\phi \mbox{ :}\varphi^2\mbox{: }
+
\frac{\lambda}{2}\phi \widehat{\varphi \varphi}.
\end{equation}
Then the contractions of the set appearing in (\ref{51}) becomes
\begin{equation}
\label{55}
-\frac{\lambda}{4!} \varphi^4
+\frac{\lambda}{2}\phi\varphi^2
-\frac{\lambda}{2}\phi^2
\Rightarrow
-\frac{\lambda}{4!}\mbox{ :}\varphi^4\mbox{: },
\end{equation}
which is clear from (\ref{52}) to (\ref{54}).
In the same way
$\frac{\delta \Delta \Gamma}{\delta \phi}
\left(\frac{\varphi^2}{2}-\phi\right)$
reduces to
\begin{equation}
\label{56}
\frac{\delta \Delta \Gamma}{\delta \phi}
\left(
\frac{1}{2}
\left(
\mbox{ :}\varphi^2\mbox{: }
+\widehat{\varphi \varphi}
\right)
-\phi\right)
=
\frac{\delta \Delta \Gamma}{\delta \phi}
\frac{1}{2}
\mbox{ :}\varphi^2\mbox{: }.
\end{equation}
In this way we get a simple formula for $\Delta \Gamma[\phi]$
\begin{equation}
\label{57}
\Delta \Gamma=
\frac{1}{i}
\left\langle
e^{-\frac{i\lambda}{4!}\varphi^4
-\frac{i}{2}\frac{\delta \Delta \Gamma}{\delta \phi}
\varphi^2}
\right\rangle_{G^{(0)}}^{\rm nself.}
=
\frac{1}{i}
\left\langle
e^{-\frac{i\lambda}{4!}\varphi^4
+\frac{i}{2}
( J^{(2)}+J^{(3)}+\cdots )
\varphi^2}
\right\rangle_{G^{(0)}}^{\rm nself.}
\end{equation}
where the superscript nself.\
implies that we have to keep all possible connected
Wick contraction using the propagator
$G^{(0)}=(\Box+m^2-J^{(0)})^{-1}$
{\em excluding self-contractions} of both the 4-point
vertex and the pseudo-vertices.

 From this formula we can successively derive
$\Gamma^{(n)}$ ($n\geq 2$)
more easily than in the previous method
in which we started from algebraic inversion formula
to obtain $J^{(n)}$ first and then  get $\Gamma^{(n)}$
through integration.
This is seen as follows.
First notice that (\ref{57}) actually starts from
$\lambda^2$
because the first order of the right-hand side
of (\ref{57}), which is
$\frac{1}{i}\left\langle
-\frac{i\lambda}{4!}\varphi^4
\right\rangle_{G^{(0)}}^{\rm nself.}$
becomes zero due to the nself.~restriction.
Since $\Gamma^{(2)}$ is of order $\lambda^2$,
we get
\begin{equation}
\label{58}
\Gamma^{(2)}=
\frac{1}{i}
\left\langle
\frac{1}{2}
\left( -\frac{i\lambda}{4!}\varphi^4 \right)^2
-\frac{i}{2}J^{(2)}
\varphi^2
\right\rangle_{G^{(0)}}^{\rm nself.}.
\end{equation}
The second term on the right-hand side makes no
contribution to $\Gamma^{(2)}$, again due to the
nself.~condition, thus leading to (\ref{G2}).
In the same way $\Gamma^{(3)}$ is calculated from the
expression
\begin{equation}
\label{59}
\Gamma^{(3)}=
\frac{1}{i}
\left\langle
\frac{1}{3!}
\left( \frac{-i\lambda}{4!}\varphi^4 \right)^3
-i\frac{\lambda}{4!}\varphi^4
\left(
-\frac{i}{2}J^{(2)}
\varphi^2
\right)
-\frac{i}{2}J^{(3)}
\varphi^2
\right\rangle_{G^{(0)}}^{\rm nself.}.
\end{equation}
and we get (\ref{G3}).

This course of study can be continued (up to the desired
order) to get (\ref{G4}) and so on.
In (\ref{57}) we do not yet have the 1VI restriction
explicitly, but we can see that, due
to the additional vertex
$ -\frac{i}{2}\frac{\delta \Delta\Gamma}{\delta \phi} \varphi^2
= -\frac{i}{2} \Delta J\varphi^2  $,
all the 1VR structures in the diagrammatic expression of
(\ref{57}) exactly cancel out.
For example the 1VR graph of Fig.~3 (a) appearing in
(\ref{57}) for $n=5$ is canceled by those of Fig.~3 (b) and (c),
which are supplied by the pseudo-vertex
$ -\frac{i}{2} \Delta J\varphi^2.  $
Thereby a practical formula for $\Delta \Gamma[\phi]$
is obtained;
\begin{description}
\item[Proposition A2)]
$\Delta \Gamma$ is given by the following rule;
\begin{equation}
\label{GAMa}
\Delta \Gamma
=
\frac{1}{i} \left\langle
e^{-\frac{i\lambda}{4!}\varphi^4
+\frac{i}{2}(J^{(2)}+ J^{(3)}+\cdots)
\varphi^2
}
\right\rangle_{G^{(0)}}^{\rm 1VI/nself.},
\end{equation}
that is, the sum of
all the connected 1VI/nself.\ vacuum diagram built
with 4-point vertices of $-\lambda$, 2-point pseudo-vertices of
$J^{(i)}$ ($i \geq 2$) and propagators $G^{(0)}$.
\end{description}
The condition 1VI/nself.\ implies that only the connected Wick
contraction corresponding to the 1VI graph need to
be considered and, at the same time, that the self-contractions
of the pseudo-vertex of Fig.~1 are excluded.
The self-contractions of 4-point vertex of Fig.~2
are automatically excluded by the restriction of 1VI.
Corresponding to the relation (\ref{32}) or
\begin{equation}
-
\setlength{\unitlength}{1mm}
\begin{picture}(18,10)(-2,0)
\thicklines
\put(4,1){\circle{8}}
\put(-0.25,1){\circle*{1}}
\put(8.25,1){\circle*{1}}
\put(12.3,1){\makebox(0,0){$\Delta J$}}
\end{picture}
=
\frac{\delta \Delta\Gamma}{\delta J^{(0)}},
\end{equation}
the ndself.~restriction in (\ref{J}) is changed to nself.\
in (\ref{GAMa}).
Proposition A$1'$) is the derivative form of Proposition A2).
Proposition A$2$)
is clearly equivalent to the following Proposition A$2'$) and
is justified rigorously in the next subsection.
\begin{description}
\item[Proposition A$\bf 2'$)]
$\Gamma^{(n)}$ ($n\geq 2$) is the sum of all possible
$n$-th order 1VI/nself.\ diagram constructed out of the 4-point
vertex of order $\lambda$ and {\em the pseudo-vertices}
 of order
$\lambda^i$ ($2\leq i < n-2$), which is  denoted by
\setlength{\unitlength}{1 mm}
\begin{picture}(25,10)
\thicklines
\put(0,0){\line(1,0){20}}
\put(10,0){\circle*{1}}
\put(10,1){\makebox(0,0)[b]{$J^{(i)}$}}
\end{picture}
and the propagator
$G^{(0)}=(\Box+m^2-J^{(0)})^{-1}$.
\end{description}
We put stress on the fact that Proposition A2) or
A$2'$) makes it
possible to write down $\Gamma^{(n)}$ ($n \geq 2$)
successively with its $\phi$ dependence coming only
through $J^{(0)}[\phi]$, although
the rule contains $J^{(2)}$,$J^{(3)}$,$J^{(4)}$,\ldots.
This is because the graphs of
$\Gamma^{(n)}$
contain $J^{(i)}$ with $i\leq n-2$ while
the graphical rule for these $J^{(i)}$ in terms of
$G^{(0)}$ propagators are already known
in (\ref{J}) or
through
$\Gamma^{(i)}$  by the relation (\ref{30});
\begin{equation}
J^{(i)}=-
\left(
\setlength{\unitlength}{1mm}
\begin{picture}(12,4)
\thicklines
\put(6,1){\circle{8}}
\put(1.75,1){\circle*{1}}
\put(10.25,1){\circle*{1}}
\put(12.5,4){\makebox(0,0)[l]{$\scriptsize -\! 1$}}
\end{picture}
\right)
\:\;\;
\frac{\delta \Gamma^{(i)}}{\delta J^{(0)}}.
\label{39}
\end{equation}
Combined with the fact that $\Gamma^{(0)}$
and $\Gamma^{(1)}$ are also given only through
$J^{(0)}$, which is clear from (\ref{GAM0}) and (\ref{GAM1})
with (\ref{0th}), $\Gamma$ itself is given by $J^{(0)}$.

 From Proposition A2) or A$2'$) we can directly obtain
(\ref{G2}) to (\ref{G4}) and
\begin{equation}
\Gamma^{(5)}=
\setlength{\unitlength}{1pt}
\begin{picture}(65,24)(0,-3)
\thicklines
\put(17,0){\circle{24}}
\put(29,0){\circle{24}}
\put(41,0){\circle*{3}}
\put(42,0){\makebox(0,0)[l]{$J^{(3)}$}}
\end{picture}
+
\setlength{\unitlength}{1pt}
\begin{picture}(55,24)(-18,-3)
\thicklines
\put(0,0){\circle{24}}
\put(-12,0){\line(5,3){18.174}}
\put(-12,0){\line(5,-3){18.174}}
\put(6.174,10.290){\line(0,-1){20.58}}
\put(12,0){\circle*{3}}
\put(13,0){\makebox(0,0)[l]{$J^{(2)}$}}
\end{picture}
+
\setlength{\unitlength}{1pt}
\begin{picture}(37,24)(0,-3)
\thicklines
\put(17,0){\circle{24}}
\put(17,12){\line(0,-1){24}}
\put(5,0){\line(1,1){12}}
\put(5,0){\line(1,-1){12}}
\put(23,0){\circle{12}}
\end{picture}
+
\setlength{\unitlength}{1pt}
\begin{picture}(40,24)(-18,-3)
\thicklines
\put(0,0){\circle{24}}
\put(-12,0){\line(5,3){18.174}}
\put(-12,0){\line(5,-3){18.174}}
\put(6.174,10.290){\line(0,-1){20.58}}
\put(12,0){\circle{8}}
\end{picture}
+\cdots+
\setlength{\unitlength}{1pt}
\begin{picture}(66,24)(-33,15)
\thicklines
\put(0,18){\oval(58,36)}
\put(-12,24){\line(1,0){24}}
\put(-12,24){\line(0,-1){24}}
\put(12,24){\line(0,-1){24}}
\put(-12,24){\line(1,-1){10}}
\put(12,0){\line(-1,1){10}}
\put(12,24){\line(-1,-1){24}}
\put(-12,24){\line(1,1){12}}
\put(12,24){\line(-1,1){12}}
\end{picture}
+\cdots
\label{G5}
\end{equation}
and so on.
The directness comes from the 1VI restriction.

Now it is convenient to introduce
the whole class of the 1VI vacuum graph ${\cal K}[A]$;
\begin{equation}
{\cal K}[A]=\left\langle
e^{-\frac{i\lambda}{4!}\varphi^4}
\right\rangle_A^{\rm 1VI}
\label{skeleton}
\end{equation}
where the propagator used in the diagram is $A$.
Note that the trivial skeleton $-\frac{1}{2i}{\rm Tr} \ln A^{-1}$
is not contained in ${\cal K}[A]$ by the definition (\ref{O}).
Thus this quantity is described as
the whole class of vacuum skeleton minus the trivial skeleton.
The whole class of the vacuum skeleton is given by
\begin{equation}
\bar{\cal K}[A]=
{\cal K}[A]
-\frac{1}{2i}{\rm Tr} \ln A^{-1}
=\int {\cal D}\varphi \left.
e^{-\frac{1}{2}\varphi A^{-1} \varphi
-\frac{i\lambda}{4!}\varphi^4}
\right|_{\rm excl. \; 1VR}
\end{equation}
where excl.~1VR implies that the 1VR graphs are excluded
or that only the 1VI graph and the trivial skeleton are kept.

In (\ref{G4}), (\ref{G5}) and in the graphs
of $\Gamma^{(n)}$ with higher $n$ obtained
by Proposition A$2'$) we see that $\Gamma^{(n)}$
is the sum of all the 1VI vacuum diagrams built with the 4-point
vertex and the decorated $G^{(0)}$ propagator.
The decoration is done by $J^{(n)}$ ($n\geq 2$)
pseudo-vertices which are inserted into the $G^{(0)}$ propagators
in all possible ways.
We see also that
$-\frac{1}{2i}{\rm Tr} \ln [G^{(0)}]^{-1}$
and the self-contractions of the pseudo-vertex
$J^{(i)}$ with $i\geq 2$ are not included in
$\Delta \Gamma$.
Thereby we arrive at Proposition A$2''$) below.

\begin{description}
\item[Proposition A$\bf 2''$)]
$\Delta \Gamma[\phi]$ is given by
${\cal K} [\bar{G}]
-\frac{1}{2i}{\rm Tr} \ln [\bar{G}]^{-1}
-\Delta {\cal K}_{\rm tr.}
=
\bar{\cal K} [\bar{G}]
-\Delta {\cal K}_{\rm tr.}
$ where
\begin{eqnarray}
\bar{G}
&=&
\left(\Box+m^2-J^{(0)}-J^{(2)}-J^{(3)}-\cdots\right)^{-1}
\label{37} \\
&=&
\left(\Box+m^2+\lambda \phi-J[\phi]\right)^{-1}
\label{37'}
\end{eqnarray}
or, with the line representing the propagator $G^{(0)}$,
\begin{equation}
\bar{G}=
\setlength{\unitlength}{1mm}
\begin{picture}(15,13)(-1,-1)
\thicklines
\put(0,0){\line(1,0){10}}
\put(15,0){\makebox(0,0){+}}
\put(20,0){\line(1,0){15}}
\put(27.5,0){\circle*{1}}
\put(27.5,3){\makebox(0,0){$J^{(2)}$}}
\put(40,0){\makebox(0,0){+}}
\put(45,0){\line(1,0){15}}
\put(52.5,0){\circle*{1}}
\put(52.5,3){\makebox(0,0){$J^{(3)}$}}
\put(70,0){\makebox(0,0){$+\cdots +$}}
\put(80,0){\line(1,0){25}}
\put(88,0){\circle*{1}}
\put(97,0){\circle*{1}}
\put(88,3){\makebox(0,0){$J^{(2)}$}}
\put(97,3){\makebox(0,0){$J^{(3)}$}}
\put(113,0){\makebox(0,0){$+\cdots$}}
\end{picture}
\label{38}
\end{equation}
and
\begin{equation}
\Delta {\cal K}_{\rm tr.}
=
-\frac{1}{2i}{\rm Tr}\ln [G^{(0)}]^{-1}
+\phi \left(J-J^{(0)}-J^{(1)}\right).
\label{e1}
\end{equation}
In other words, $\Gamma=\Gamma^{(0)}+\Gamma^{(1)}+\Delta \Gamma$
is given by,
\begin{eqnarray}
\Gamma[\phi]
&=&
-\phi J[\phi]+\frac{\lambda}{2}\phi^2
-\frac{1}{2i}{\rm Tr}\ln
\left(\Box+m^2-J[\phi]+\lambda\phi\right)
+{\cal K}[\bar{G}]
\nonumber \\
&=&
-\phi J[\phi]+\frac{\lambda}{2}\phi^2
+\bar{\cal K}[\bar{G}].
\label{2.1a}
\end{eqnarray}
\end{description}
The quantity
$
-\frac{1}{2i}{\rm Tr} \ln [\bar{G}]^{-1}
-\Delta {\cal K}_{\rm tr.}
$
is a ${\rm Tr}\ln$ of a decorated propagator specified as follows.
The decoration is made by
$J^{(0)}, J^{(2)},J^{(3)},\ldots$
($J^{(1)}=\lambda\phi$ is not included) but
the decoration only by $J^{(0)}$'s
(the first term on the right-hand side of  (\ref{e1}))
and the decoration by one single
$J^{(2)},J^{(3)},\ldots$
(the last term in (\ref{e1}),
which is a summation of the self-contracted diagrams of Fig.~1
with $i\geq 2$)
are excluded.
Proposition A$2''$)  will be justified
in the next subsection precisely.

Although the appearance of the term
$-J\phi$ in (\ref{2.1a}) seems to be somewhat
curious it is not actually so.
Differentiating (\ref{2.1a}) with respect to
$\phi$ by noting (\ref{37'}) and (\ref{8})
we get
\begin{equation}
\left(
\phi+\frac{\delta \bar{\cal K}}{\delta \bar{G}^{-1}}
\right)
\left(
-\frac{\delta J}{\delta \phi}+\lambda
\right)
=0.
\end{equation}
The second term is not zero because
$\frac{\delta J}{\delta \phi}$
contains various orders of $\lambda$.
Thus we get
\begin{equation}
\phi=-\frac{\delta \bar{\cal K}}{\delta \bar{G}^{-1}},
\label{e2}
\end{equation}
which is consistent with (\ref{6})
or
$\phi=\frac{\delta W}{\delta J}$
in the following sense.
If one uses the relation
\begin{equation}
W=\Gamma+J\phi=
\frac{\lambda}{2}\phi^2+\bar{\cal K}[\bar{G}]
\end{equation}
(obtained from (\ref{2.1a}))
in the right-hand side of
$\phi=\frac{\delta W}{\delta J}$
and then uses (\ref{e2})
one gets the left-hand side of this equation, that is, $\phi$.
\subsubsection{Rule for $\Delta \Gamma$ in terms of $J^{(0)}[\phi]$}
 From Proposition A2) and A$1'$) we can deduce another
graphical rule for $\Delta \Gamma$ and $\Delta J$
in which all the $\phi$ dependence is explicitly through
$J^{(0)}[\phi]$.
We arrive at the new rule by using (\ref{39}) and
by noting that $\frac{\delta \Gamma^{(i)}}{\delta J^{(0)}}$
is given by the right-hand side of (\ref{J})
(in addition to the facts stated just above Proposition A$2''$)).
To state the new rule we introduce $i$-vertex
($i=0,1,2,\ldots$), which is defined as
\begin{equation}
v_i(x_1,\ldots,x_i)
\equiv
\frac{1}{i!}
\frac{\delta^i \bar{\cal K}[G^{(0)}]}{
\delta J^{(0)}(x_1) \cdots \delta J^{(0)}(x_i)}
-\delta_{i,1}
\setlength{\unitlength}{1mm}
\begin{picture}(10,10)(-2,0)
\thicklines
\put(4,1){\circle{8}}
\put(-0.25,1){\circle*{1}}
\end{picture}
-\delta_{i,2}
\setlength{\unitlength}{1mm}
\begin{picture}(10,10)(-2,0)
\thicklines
\put(4,1){\circle{8}}
\put(-0.25,1){\circle*{1}}
\put(8.25,1){\circle*{1}}
\put(10,0){\makebox(0,0){.}}
\end{picture}
\label{41}
\end{equation}
where
$\delta_{i,j}$ is the Kronecker delta.

Now the final rule is given by the following
statement where {\em the graphs
are built with the inverse composite propagator $D(x,y)$ and the
vertices $v_i(x_1,\ldots,x_i)$} ($i=0,1,2,\ldots$).
\begin{description}
\item[Proposition A$\bf 3$)]
$\Delta \Gamma$ and $\Delta J$
is given by the following rule;
\begin{eqnarray}
\Delta \Gamma &=& \sum \left[ \right.
\mbox{ all the connected {\em tree} diagrams
with all the pairs of the argument }
\nonumber \\
& &
\mbox{ of $v_i$'s connected by $D$ propagators
(vacuum graph).}
\left. \right]
\label{43'}
\end{eqnarray}
\begin{eqnarray}
\Delta J(x)
&=& \int d^4 y D_{xy} \times \sum \left[ \right.
\mbox{ all the connected {\em tree} diagrams
with one of the argument }
\nonumber \\
& &
\mbox{ of one of the
$v_i$'s being the point $y$
(graph with an external point).}
\left. \right]
\end{eqnarray}
\end{description}
The tree graph in terms of $D$ propagator is the graph
in which all the $D$ propagators are {\em articulate}.
Here the $D$ propagator in a connected graph is called
{\em articulate} if removal of it leads to a separation
of the graph.
Note that $D(x,y)$ lines never make a loop because
$D(x,y)$ comes from $J^{(i)}$ with $i\geq 2$
(see (\ref{39})).
Proposition A3) is understood by examples.
For instance,
$\Gamma^{(4)}$ in (\ref{G4}) or
\begin{equation}
\label{G4'}
\Gamma^{(4)}=
-
\setlength{\unitlength}{1mm}
\begin{picture}(15,10)
\thicklines
\put(5,1){\circle{8}}
\put(13.25,1){\circle*{1}}
\put(9,1){\circle{8}}
\end{picture}
\left(
\setlength{\unitlength}{1mm}
\begin{picture}(12,4)
\thicklines
\put(6,1){\circle{8}}
\put(1.75,1){\circle*{1}}
\put(10.25,1){\circle*{1}}
\put(12.5,4){\makebox(0,0)[l]{$\scriptsize -\! 1$}}
\end{picture}
\right)
\setlength{\unitlength}{1mm}
\begin{picture}(18,10)(-2,0)
\thicklines
\put(5,1){\circle{8}}
\put(0.75,1){\circle*{1}}
\put(9,1){\circle{8}}
\end{picture}
+
\setlength{\unitlength}{1pt}
\begin{picture}(49,24)(0,-3)
\thicklines
\put(17,0){\circle{24}}
\put(29,0){\circle{24}}
\put(35,0){\circle{12}}
\end{picture}
+
\setlength{\unitlength}{1pt}
\begin{picture}(51,24)(5,-3)
\thicklines
\put(17,0){\circle{16}}
\put(29,0){\circle{24}}
\put(41,0){\circle{16}}
\end{picture}
+
\setlength{\unitlength}{1pt}
\begin{picture}(40,24)(-18,-3)
\thicklines
\put(0,0){\circle{24}}
\put(8.485,8.485){\line(0,-1){16.97}}
\put(8.485,8.485){\line(-1,0){16.97}}
\put(-8.485,-8.485){\line(0,1){16.97}}
\put(-8.485,-8.485){\line(1,0){16.97}}
\end{picture}
\end{equation}
can be written as
\begin{equation}
\Gamma^{(4)}=\mbox{the fourth order of}
\quad
\left(
\frac{1}{2}v_1 D v_1 + v_0
\right).
\end{equation}
In (\ref{G4'}) $J^{(0)}[\phi]$ dependence is evident because there is
no $J^{(i)}$ pseudo-vertex ($i \geq 2$).
All the $\phi$-dependence is through $J^{(0)}$ contained in
$G^{(0)}$ (and $D$).
The sum of the first two terms of (\ref{G4}) exactly coincide with the
first term of (\ref{G4'}) with correct weight
after substitution of (\ref{J2}) or (\ref{j2}).

Proposition A3) can be expressed as follows.
For this purpose we introduce $\sigma$-field which has
the propagator $D$. (The $\sigma$-field looks like the
auxiliary field but has nothing to do with it.)
Then $\Delta \Gamma$ is given by
\begin{eqnarray}
\Delta \Gamma
&=&
\frac{1}{i}
\left.
\frac{\int {\cal D}\varphi{\cal D}\sigma
e^{iS_0}e^{iS_{int.}}}{\int {\cal D}\varphi{\cal D}\sigma
e^{iS_0}}
\right|_{\rm conn./ tree /1VI/excl.}
\nonumber \\
S_0
&=&
-\frac{1}{2}\varphi \left[G^{(0)}\right]^{-1} \varphi
-\frac{1}{2}\sigma D^{-1} \sigma
\nonumber \\
S_{int.}
&=&
-\frac{\lambda}{4!}\varphi^4
+\frac{1}{2}\sigma\varphi^2
\label{GAMb}
\end{eqnarray}
with
\begin{equation}
D=
-\frac{\delta J^{(0)}}{\delta \phi}
=-\left(
\setlength{\unitlength}{1mm}
\begin{picture}(12,4)
\thicklines
\put(6,1){\circle{8}}
\put(1.75,1){\circle*{1}}
\put(10.25,1){\circle*{1}}
\put(13,4){\makebox(0,0)[l]{$\scriptsize -1$}}
\end{picture}
\right).
\end{equation}
Here the subscript conn./tree/1VI/excl.\
implies that only  {\em connected} graphs
which is {\em tree} in terms of $D$ propagator and also
{\em 1VI}\/ in terms of the 4-point vertex have to be considered
and, at the same time, that the sub-structure of
\setlength{\unitlength}{1mm}
\begin{picture}(10,5)
\thicklines
\put(5,1){\circle{8}}
\put(0.75,1){\circle*{1}}
\end{picture}
and
\setlength{\unitlength}{1mm}
\begin{picture}(12,5)
\thicklines
\put(5,1){\circle{8}}
\put(0.75,1){\circle*{1}}
\put(9.25,1){\circle*{1}}
\end{picture}
has to be excluded.
Hence with the compact and self-evident notation,
Proposition A3) is rewritten in the following form;
\begin{description}
\item[Proposition A$\bf 3'$)]
$\Delta \Gamma$ and $\Delta J$
is given by the following formula;
\begin{equation}
\label{GAMc}
\Delta \Gamma
=
\frac{1}{i}
\left\langle
e^{-\frac{i\lambda}{4!}\varphi^4+\frac{i}{2}\sigma\varphi^2
}
\right\rangle_{G^{(0)},D}^{\rm tree/1VI/excl.}
\end{equation}
\begin{equation}
\label{DELJ}
-
\setlength{\unitlength}{1mm}
\begin{picture}(20,6)(0,2)
\thicklines
\put(6,3){\circle{10}}
\put(1,3){\circle*{1}}
\put(11,3){\circle*{1}}
\put(15,3){\makebox(0,0){$\Delta J $}}
\end{picture}
=
\left\langle
\frac{1}{2}\varphi^2
e^{-\frac{i\lambda}{4!}\varphi^4+\frac{i}{2}\sigma\varphi^2
}
\right\rangle_{G^{(0)},D}^{\rm tree/1VI/excl.}
\end{equation}
where the connected graphs with tree/1VI/excl.~restriction
are constructed by  3-point ($\sigma\phi^2$) and 4-point
($\lambda\phi^4$) vertices and propagators $G^{(0)}$ of the $\varphi$
field and $D$ of the $\sigma$ field.
\end{description}
Recall that both $G^{(0)}$ and $D$ are functionals of $J^{(0)}$.
Proposition A$3'$) is easily understood from the rule (\ref{GAMa})
with (\ref{J})
but a rigorous proof is presented in Sec.~2.2.
Notice that this rule does not contain the $J^{(i)}$
pseudo-vertex unlike the previous rules but instead
$D$ is represented by the propagator of the artificially introduced
$\sigma$ field.
 From (\ref{GAMc}), the quantity $\Gamma^{(4)}$, for example, can be
{\em directly} obtained as (\ref{G4'}) above.

Finally we note that a certain infinite series of the
graphs appearing in $\Delta \Gamma [\phi]$ can be
conveniently summed up. The series $\Gamma_{ch.}$ is
the sum of all the possible closed chains constructed
out of the unit element
\setlength{\unitlength}{1mm}
\begin{picture}(10,10)
\thicklines
\put(5,1){\circle{8}}
\put(0.75,1){\circle*{1}}
\put(9.25,1){\circle*{1}}
\end{picture}
or
\begin{equation}
\Gamma_{ch.}=
\setlength{\unitlength}{1pt}
\begin{picture}(50,34)(0,-3)
\thicklines
\put(17,0){\circle{24}}
\put(29,0){\circle{24}}
\end{picture}
+
\setlength{\unitlength}{1pt}
\begin{picture}(40,24)(-18,-3)
\thicklines
\put(0,0){\circle{24}}
\put(12,0){\line(-5,3){18.174}}
\put(12,0){\line(-5,-3){18.174}}
\put(-6.174,10.290){\line(0,-1){20.58}}
\end{picture}
+
\setlength{\unitlength}{1pt}
\begin{picture}(40,24)(-18,-3)
\thicklines
\put(0,0){\circle{24}}
\put(8.485,8.485){\line(0,-1){16.97}}
\put(8.485,8.485){\line(-1,0){16.97}}
\put(-8.485,-8.485){\line(0,1){16.97}}
\put(-8.485,-8.485){\line(1,0){16.97}}
\end{picture}
+\cdots.
\label{44}
\end{equation}
This series is summed up to give
\vspace{1mm}
\begin{equation}
\Gamma_{ch.}=
-\frac{1}{2}\left[\:
{\rm Tr}\ln\left(1-\lambda
\setlength{\unitlength}{1mm}
\begin{picture}(10,5)
\thicklines
\put(5,1){\circle{8}}
\put(0.75,1){\circle*{1}}
\put(9.25,1){\circle*{1}}
\end{picture}
\right)
+\lambda {\rm Tr}
\setlength{\unitlength}{1mm}
\begin{picture}(10,5)
\thicklines
\put(5,1){\circle{8}}
\put(0.75,1){\circle*{1}}
\put(9.25,1){\circle*{1}}
\end{picture}
+(3!-2)
\setlength{\unitlength}{1mm}
\begin{picture}(15,10)
\thicklines
\put(5,1){\circle{8}}
\put(9,1){\circle{8}}
\end{picture}
\right].
\label{chain}
\end{equation}
\subsection{Formal justification of Propositions}
In this subsection we directly prove
Proposition A$2''$), A$1'$), and A$3'$), leading to the
full proof of all the propositions in the previous
subsection.

Proposition A$2''$) or (\ref{2.1a}) is proved easily
by analyzing the graphic expression of $W[J]$ rather
than that of $\Gamma[\phi]$.
It is based on a similar topological proof given
in \cite{DM}.
If one writes the graph rule of $W[J]$ using
$G_J=(\Box+m^2-J)^{-1}$ as the propagator (the
rule (\ref{W1})),
the whole dependence of $W[J]$ on $J$ is through the propagator
$G_J=(\Box+m^2-J)^{-1}$.
The contribution of all the 1VI graphs appearing in $W[J]$
can be written as ${\cal K}[G_J]$, the vacuum skeleton minus
the trivial skeleton
$-{\rm Tr}\ln G_J^{-1}$ (see (\ref{skeleton})).
Then all the graphs of $W[J]$ seem to be generated by
replacing $G_J$ with $[G_{J}^{-1}-(-\lambda\phi)]^{-1}$,
that is, $W[J]$ seems to be given by
\begin{equation}
-\frac{1}{2i}{\rm Tr} \ln \left (G^{-1}_J+\lambda \phi \right)
+{\cal K}[(G^{-1}_J+\lambda\phi)^{-1}]=\bar{\cal K}[\bar{G}].
\label{ee1}
\end{equation}
Note here that $\phi$ is the sum of the all distinct connected
diagrams with one external point where two propagators meet.
(We here use the rule (\ref{W1}) and (\ref{PHI1})
in which the propagator
$G_J=(\Box+m^2-J)^{-1}$ is used
so that there are only the
4-point vertices $-\lambda$ and the pseudo-vertex does not
exist in the graphs of $\phi$.)
But the above statement is not exactly true because each element
of the graphs of (\ref{ee1}) is incorrectly weighted.
To examine this point
the number of the skeletons $N(\bar{\cal K})$ is defined as follows.
Removal of all 1VR vertices in a graph leads to separated
graphs which no longer have any lines connecting them.
Then all the resulting separated graphs are skeletons and the
number of them is $N(\bar{\cal K})$.
Note here that the skeleton and $v_j$ vertex are slightly different,
that is, $v_j$ does not contain
the second and the last term in (\ref{41}) and the trivial skeleton
while the skeleton does.
An example of the graph of $N(\bar{\cal K})=4$ is given
in Fig.~4.
Now we see that each graph of $W[J]$ is contained
in (\ref{ee1}) $N(\bar{\cal K})$ times.

On the other hand, if we turn our attention to 1VR vertices
the graphs of $W[J]$ seem to be generated
by
\begin{equation}
\frac{1}{2}\phi(-\lambda)\phi=-\frac{\lambda}{2}\phi^2
\label{ee2}
\end{equation}
because $\phi$ is all the distinct connected graphs with one
external point (given by the rule (\ref{PHI1})).
Again this is not true however because each element of $W[J]$
appears $N$(1VR) times where $N$(1VR) is the number
of 1VR vertices in the graph.

Thus the above two ways to construct the graphs of $W[J]$
is not satisfactory. But fortunately we have a simple
topological relation
\begin{equation}
N(\bar{\cal K})-N(\mbox{1VR})=1.
\end{equation}
This can be proved by noting that the addition of one
skeleton having one external point
necessarily increases the number of the 1VR vertex by one.
Thus if we take the sum
$\bar{\cal K}[\bar{G}]+\frac{\lambda}{2}\phi^2$,
each graph of $W[J]$ is contained exactly once or with
correct weight.
Hence we have
\begin{equation}
W[J]= \bar{\cal K}[\bar{G}]
+\frac{\lambda}{2}\phi^2.
\end{equation}
This proves (\ref{2.1a}) or Propositions A$2''$),
leading to the proof of Propositions A2) and A$2'$).

We show below that Proposition A$2''$) can also be
proved by use of the sum-up rule, which is established by
the author\cite{KOBS,REV}.
Indeed we see that eq.~(\ref{e2}) is directly
obtained by the sum-up rule in the following.
If eq.~(\ref{e2}) holds, by assuming the form
\begin{equation}
\Gamma=-J[\phi]\phi+\frac{\lambda}{2}\phi^2+\Delta[\bar{G}],
\end{equation}
we immediately know by differentiation with respect to
$\phi$ that $\Delta[\bar{G}]$ is equal to $\bar{\cal K}[\bar{G}]$,
leading to  (\ref{2.1a}).

In order to prove (\ref{e2}) first we note that $\phi$ is all the
distinct graphs with one external point
(representing the insertion of $\varphi(x)^2$)
which are built
with the propagators $G_J=(\Box+m^2-J)^{-1}$ and the 4-point
vertices $-\lambda$ (the rule (\ref{PHI1})).
A 1-part is a subdiagram connected to the rest by one 4-point
vertex.
When cut out, the 1-part itself is one element of the graphs of $\phi$
(see Fig.~5).
The sum-up rule is best explained by an example.
In short it guarantees that we can sum up the
graphs on the left-hand side of the following example
to the single graph on the right-hand side
{\em with correct weight}.
\begin{equation}
\setlength{\unitlength}{1pt}
\begin{picture}(50,40)(0,-3)
\thicklines
\put(0,0){\circle*{3}}
\put(12,0){\circle{24}}
\put(24,0){\circle{24}}
\put(38.14,14.14){\circle{16}}
\put(38.14,-14.14){\circle{16}}
\end{picture}
+
\setlength{\unitlength}{1pt}
\begin{picture}(60,40)(-6,-3)
\thicklines
\put(0,0){\circle*{3}}
\put(12,0){\circle{24}}
\put(24,0){\circle{24}}
\put(38.14,14.14){\circle{16}}
\put(43.8,19.8){\circle{16}}
\put(38.14,-14.14){\circle{16}}
\end{picture}
+\cdots+
\setlength{\unitlength}{1pt}
\begin{picture}(80,40)(-10,-3)
\thicklines
\put(0,0){\circle*{3}}
\put(12,0){\circle{24}}
\put(24,0){\circle{24}}
\put(38.14,14.14){\circle{16}}
\put(43.8,19.8){\circle{16}}
\put(43.8,31.8){\circle{8}}
\put(55.8,19.8){\circle{8}}
\put(38.14,-14.14){\circle{16}}
\put(43.8,-19.8){\circle{16}}
\put(55.11,-31.11){\circle{16}}
\end{picture}
+\cdots=
\begin{picture}(80,40)(-7,-3)
\thicklines
\put(0,0){\circle*{3}}
\put(12,0){\circle{24}}
\put(24,0){\circle{24}}
\put(32.49,8.49){\circle*{3}}
\put(32.49,-8.49){\circle*{3}}
\put(49,10){\makebox(0,0){$\footnotesize (\!-\lambda\phi\!)$}}
\put(49,-10){\makebox(0,0){$\footnotesize (\!-\lambda\phi\!)$}}
\end{picture}.
\end{equation}
\vspace*{20pt}
In other words {\em all the 1-parts directly attached
to the skeleton through an external point are summed up to
$\phi$}.
The statement is proved rigorously as follows.

In the graphs of $\phi$, we can easily show that {\em if two different
1-parts have a common part, one completely contains the
other.}
(Note here that in a vacuum graph this is not true
so that the following arguments can not be applied to the graphs
with no external point.)
Thus one can {\em unambiguously} proceed to a larger 1-part starting
from one of the 1-parts (which is smaller) in the graph
and finally reach {\em the second largest 1-part}.
See Fig.~6 as an example.
(The largest 1-part is the whole graph itself.)
This procedure can be repeated to reach the second largest
1-part starting from another 1-part which is not contained
in the former second largest 1-parts.
We continue this until there are no 1-parts other than the second
largest ones.
Thereby we find the second largest 1-part structure of
the graph.
This operation to find the 1-part
structure is done for all the graphs of $\phi$.
After the operation we sum up
all the graphs having the same structure.
We thus know {\em all the propagators in the graphs are modified
to $\bar{G}=(G_J^{-1}+\lambda\phi)^{-1}$ while 1VR graphs disappear}
because all the second largest 1-parts are summed up to
$\phi$ with correct weight.

Hence we know that $\phi$ is all the distinct 1VI graph
(including a derivative of trivial skeleton) with one
external point where propagator
$G_J=(\Box+m^2-J)^{-1}$ is replaced by $\bar{G}$ or
\begin{equation}
\phi=\left.\frac{\delta \bar{\cal K}[G]}{\delta J}
\right|_{J\rightarrow J-\lambda\phi},
\end{equation}
which is equivalent to (\ref{e2}).
Thus (\ref{2.1a}) or Proposition A$2''$) is justified.

Having shown that Propositions A2), A$2'$) and A$2''$) are true
we can take it for granted that Propositions A$1$)
and A$1'$) also hold because Proposition A$1'$) can be
regarded as the derivative form of Proposition A2).
But Proposition A1) or A$1'$)
 can be directly proved by using the
sum-up rule again.
 From the rule (\ref{PHI2}) or (\ref{A}) we know
\begin{equation}
-
\setlength{\unitlength}{1mm}
\begin{picture}(25,5)(0,9)
\thicklines
\put(10,10){\circle{10}}
\put(5,10){\circle*{1}}
\put(15,10){\circle*{1}}
\put(20,10){\makebox(0,0){$\Delta J$}}
\end{picture}
=
\left\langle
\frac{1}{2}\varphi^2
e^{-\frac{i\lambda}{4!}\varphi^4+
\frac{i}{2}(
J^{(1)}+ J^{(2)}+\cdots)\varphi^2 }
\right\rangle_{G^{(0)}}^{\rm excl.}
\end{equation}
where the superscript {\em excl.} means that the contributions of the
0-th order and the first order in $\lambda$ and the derivative
of the self-contractions
of $J^{(i)}$ with $i\geq 2$ are {\em excluded}\/ from the expression.
The derivative
of the self-contractions have been moved on the left-hand side.
Keeping the graphical meaning of
(\ref{PHI2}) in mind we apply the sum-up rule again
to obtain
\begin{eqnarray}
-
\setlength{\unitlength}{1mm}
\begin{picture}(25,5)(0,9)
\thicklines
\put(10,10){\circle{10}}
\put(5,10){\circle*{1}}
\put(15,10){\circle*{1}}
\put(20,10){\makebox(0,0){$\Delta J$}}
\end{picture}
&=&
\mbox{all distinct {\em 1VI graphs with one external point}}
\nonumber \\
&&
\mbox{which are built with the propagator}
\nonumber \\
&&
\mbox{$\left(\left[G^{(0)}\right]^{-1}
+\lambda\phi\right)^{-1}$ and the 4-point vertex $-\lambda$}
\nonumber \\
&&
\mbox{and the pseudo-vertex $J^{(i)}$ with $i\geq 1$.}
\nonumber \\
&&
\end{eqnarray}
In the above, all the corrections by the pseudo-vertex
$J^{(1)}\varphi^2$ change the propagator
$\left(\left[G^{(0)}\right]^{-1}+\lambda\phi\right)^{-1}$
back to $G^{(0)}$ hence we get
\begin{equation}
-
\setlength{\unitlength}{1mm}
\begin{picture}(25,5)(0,9)
\thicklines
\put(10,10){\circle{10}}
\put(5,10){\circle*{1}}
\put(15,10){\circle*{1}}
\put(20,10){\makebox(0,0){$\Delta J$}}
\end{picture}
=
\left\langle
\frac{1}{2}\varphi^2
e^{-\frac{i\lambda}{4!}\varphi^4+
\frac{i}{2}(
J^{(2)}+ J^{(3)}+\cdots)\varphi^2 }
\right\rangle_{G^{(0)}}^{\rm 1VI/ndself.}
=
\left\langle
\frac{1}{2}\varphi^2
e^{-\frac{i\lambda}{4!}\varphi^4+
\frac{i}{2}\Delta J
\varphi^2 }
\right\rangle_{G^{(0)}}^{\rm 1VI/ndself.}.
\label{eqeq}
\end{equation}
This equation is, of course, equivalent to
Proposition A$1'$).

The remaining work is to prove
Proposition A$3'$). First the rule (\ref{DELJ}) for $\Delta J$ is
easily proved by mathematical induction;
we assume the rule is true up to $J^{(n)}$ or the $n$-th order of
$\Delta J$ and then we can convince
ourselves that the statement for $J^{(n+1)}$ or the ($n+1$)-th order
of $\Delta J$ is also true from Proposition A$1'$).
For this purpose we have only to note
that the graphs of $J^{(n)}$ contain $J^{(i)}$
($i\leq n-2$)  and have one external point so that
the sum-up rule can be applied.

The last task is to prove\footnote{
The author got the idea of the proof presented below
from S. Yokojima, to whom he is very thankful.}
the rule (\ref{GAMc}) for $\Delta \Gamma$.
It is clear from Propositions A2) and A$1'$)
that all the graphs appearing in $\Delta \Gamma$
are exhausted in the rule (\ref{GAMc}).
Thus it is enough if
we confirm that the graphs  of $\Delta \Gamma$
in Proposition A$2''$) appear with the same weight
as in the rule (\ref{GAMc}).
In other words we justify (\ref{GAMc}) on the basis of
Proposition A$2''$).
To this end, we expand
$-\frac{1}{2i}{\rm Tr} \ln [\bar{G}]^{-1}$
in terms of $\Delta J$ ($=J^{(2)}+J^{(3)}+\cdots$)
and get
\begin{equation}
-\frac{1}{2i}{\rm Tr} \ln [\bar{G}]^{-1}
-\Delta {\cal K}_{\rm tr.}
=\sum_{n=2}^{\infty}
\frac{1}{2in}{\rm Tr}(G^{(0)}\Delta J)^n.
\end{equation}
$\Delta {\cal K}_{\rm tr.}$ is canceled by the 0-th and first order
of the expansion.
Therefore we get,
from the expression $\Delta \Gamma={\cal K}[\bar{G}]
-\frac{1}{2i}{\rm Tr} \ln [\bar{G}]^{-1}
-\Delta {\cal K}_{\rm tr.}$,
\begin{eqnarray}
\Delta \Gamma
&=&
{\cal K}[\bar{G}]
+
\sum_{n=3}^{\infty}\frac{1}{2in}{\rm Tr}
\left(G^{(0)}\Delta J\right)^n
+
\setlength{\unitlength}{1mm}
\begin{picture}(25,5)(-2.5,-1)
\thicklines
\put(1,0){\makebox(0,0){$\Delta J$}}
\put(10,0){\circle{10}}
\put(5,0){\circle*{1}}
\put(15,0){\circle*{1}}
\put(19,0){\makebox(0,0){$\Delta J$}}
\end{picture}
\nonumber \\
&=&
{\cal K}\left[([G^{(0)}]^{-1}-\Delta J)^{-1}\right]
+
\sum_{n=3}^{\infty}
\setlength{\unitlength}{1mm}
\begin{picture}(25,10)(-6,-1)
\thicklines
\put(0,0){\circle{10}}
\put(3.54,3.54){\circle*{1}}
\put(3.54,-3.54){\circle*{1}}
\put(8,4.5){\makebox(0,0){$\Delta J$}}
\put(7,-4){\makebox(0,0){$\Delta J$}}
\put(2.5,-2.5){\makebox(0,0){$\scriptsize 1$}}
\put(2.5,2.5){\makebox(0,0){$\scriptsize n$}}
\put(8,1.5){\circle*{.2}}
\put(8.25,0){\circle*{.2}}
\put(8,-1.5){\circle*{.2}}
\end{picture}
\nonumber
\end{eqnarray}
\begin{equation}
\quad \quad\quad\quad
-
\frac{1}{2}
\left(
\setlength{\unitlength}{1mm}
\begin{picture}(25,6)(-4,-1)
\thicklines
\put(1,0){\makebox(0,0){$-\Delta J$}}
\put(13,0){\circle{10}}
\put(8,0){\circle*{1}}
\put(18,0){\circle*{1}}
\end{picture}
\right)
D
\left(
\setlength{\unitlength}{1mm}
\begin{picture}(25,6)(-1,-1)
\thicklines
\put(1,0){\makebox(0,0){$-$}}
\put(10,0){\circle{10}}
\put(5,0){\circle*{1}}
\put(15,0){\circle*{1}}
\put(19,0){\makebox(0,0){$\Delta J$}}
\end{picture}
\right).
\label{topo}
\end{equation}
By this relation the
rule for $\Delta \Gamma$ is also proved by mathematical
induction.
We assume that the rule is true up to the $n$-th order of
$\Delta \Gamma$ or $\Gamma^{(n)}$.
We notice here that
the first two terms on the right-hand side of
(\ref{topo}) contain each graph $N(v_j)$ times and the
last term $N(D)$ times (see the graphical
rule (\ref{DELJ}) for $\Delta J$).
Here $N(v_j)$ and $N(D)$ are the number of
$v_j$ vertices ($j=1,2,\ldots$) and that of the $D$ propagators
respectively.
Due to the topological relation
\begin{equation}
N(v_j)-N(D)=1
\end{equation}
we confirm that $\Gamma^{(n+1)}$ is given
correctly by the final rule (\ref{GAMc}).
\setcounter{equation}{0}
\renewcommand{\theequation}{\arabic{section}.\arabic{equation}}
\section{Case of itinerant electron model}
\label{s3}
In the previous section we have taken the $\varphi^4$
theory which is simple and
convenient to develop a general framework.
In this section we take a physically more interesting
system as another example --- the itinerant electron model
including the Hubbard model.
We couple an external source
to the local composite operator corresponding
to the spin operator (and to the number density operator).
Writing down the effective action for such a system is
equivalent to rewrite the theory in terms of the
expectation value of the spin operator
or the magnetizatoin instead
of the external source or the magnetic field.
Such a formulation is of course convenient for the
study of magnetic phase of the system --- problem
of the spontaneous symmetry breaking of SU(2),
which is inherent in the model.

The generating functional for this system
(written as $\Omega$ in this section instead of $W$)
is a generalization of the thermodynamical potential
to the case where an external source, {\em which depends
on imaginary time} $\tau$, is present.
This is particularly useful for our purpose and is defined by
\begin{equation}
\label{3.1}
e^{-\Omega[J]}={\rm Tr}T_\tau
e^{-\int_0^\beta d\tau {\cal H}[J]}
\end{equation}
\begin{eqnarray}
\label{3.2}
{\cal H}[J]&=&{\cal H}_0+{\cal H}_J
\\
{\cal H}_0 &=&
\sum_{\bf r r'}\sum_\sigma
t_{\bf r r'} a^\dagger_{{\bf r}\sigma}a_{{\bf r'}\sigma}
+U\sum_{\bf r}n_{{\bf r}\uparrow}n_{{\bf r}\downarrow}
\label{3.3}
\\
{\cal H}_J&=&
-\sum_{{\bf r}\sigma}J_\sigma({\bf r}\tau)n_{{\bf r}\sigma}
\label{3.4}
\\
&=&
-\sum_{\bf r} h({\bf r}\tau)\hat{S}_z({\bf r})
-\mu N
\label{3.5}
\end{eqnarray}
where $\beta^{-1}$ is the temperature of
the system and $T_\tau$ is the $\tau$-ordering operator.
The creation and annihilation operators for the electron
of spin $\sigma$ and $\sigma'$ at the lattice site
$\bf r$ and $\bf r'$ satisfy
\begin{equation}
\label{3.6}
\{a_{{\bf r}\sigma},a^\dagger_{{\bf r'}\sigma'}\}
=\delta_{\bf r r'}\delta_{\sigma \sigma'}.
\end{equation}
Furthermore $t_{\bf r r'}$ represents the hoping term
and $U$ the on-site Coulomb interaction.
We have also introduced
\begin{eqnarray}
\label{3.7}
n_{{\bf r}\sigma} &=&
a^\dagger_{{\bf r}\sigma}a_{{\bf r}\sigma},
\\
\hat{S}_z({\bf r}) &=&
\frac{1}{2}(n_{{\bf r}\uparrow}-n_{{\bf r}\downarrow}),
\label{3.8}
\\
N &=& \sum_{\bf r}
(n_{{\bf r}\uparrow}+n_{{\bf r}\downarrow}),
\label{3.9}
\\
J_\sigma({\bf r}\tau)&=&
\frac{\sigma}{2}h({\bf r}\tau)+\mu.
\label{3.10}
\end{eqnarray}
We regard below both the chemical potential and
the $\tau$-dependent magnetic field
$h({\bf r}\tau)$
as external sources
for convenience.
They are combined to $J_\sigma({\bf r}\tau)$
as in (\ref{3.10}).
Note here that if we want to rewrite the theory in terms
of the expectation value of the number density operator
without taking the spin operator
as another dynamical variable
we have only to set $J_\uparrow=
J_\downarrow$ in the following formulae.
The spin index $\sigma$ is defined to take the value
($+1$,$-1$) for ($\uparrow$,$\downarrow$).

The path-integral representation in terms of Grassmann
variables $z$ and $z^*$ (corresponding to
the operators $a$ and $a^\dagger$ respectively)
is given by (see Appendix D)
\begin{equation}
\label{3.14}
e^{-\Omega}
=
\int {\cal D}z^* {\cal D}z e^{S[z^*,z,J]}
\end{equation}
\begin{eqnarray}
S[z^*,z,J]
&=&
-\sum_{xx'\sigma}
z^*_{x\sigma} [G_{\sigma}]^{-1}_{xx'} z_{x'\sigma}
-U\sum_x
z^*_{x\uparrow}z_{x\uparrow}
z^*_{x\downarrow}z_{x\downarrow}
+\sum_{x\sigma}J_{x\sigma}
z^*_{x\sigma}z_{x\sigma}
\label{3.15}
\\
&\equiv&
-\sum_\sigma z^*_\sigma G^{-1}_{J\sigma} z_\sigma
-U
z^*_{\uparrow}z_{\uparrow}
z^*_{\downarrow}z_{\downarrow}
\label{3.16}
\\
G^{-1}_{x x'}
&=&
\delta_{\tau \tau'}
\left(
\delta_{\bf r r'}\frac{\partial}{\partial \tau'}
+t_{\bf r r'}
\right)
\quad
\mbox{and}
\quad
\left[
G^{-1}_{J\sigma}
\right]_{xx'}
=
G^{-1}_{xx'}
-
\delta_{\tau\tau'}
\delta_{\bf r r'}
J_{x\sigma}
\label{3.17}
\end{eqnarray}
where $x$ and $x'$ denote the sets (${\bf r}\tau$) and
(${\bf r'}\tau'$) respectively.
 From this expression it is straightforward to get the Feynman
diagram expansion for $\Omega$ in powers of $U$.
The expectation value of the local number operator
$n_{{\bf r}\sigma}$ is defined as
\begin{equation}
\phi_{x\sigma}=
-\frac{\delta \Omega}{\delta J_{x\sigma}}
=\langle
a^\dagger_{{\bf r}\sigma}a_{{\bf r}\sigma}
\rangle_\tau
=\left\langle
\frac{
a^\dagger_{{\bf r}\uparrow}a_{{\bf r}\uparrow}
+
a^\dagger_{{\bf r}\downarrow}a_{{\bf r}\downarrow}
}{2}
+\sigma
\frac{
a^\dagger_{{\bf r}\uparrow}a_{{\bf r}\uparrow}
-
a^\dagger_{{\bf r}\downarrow}a_{{\bf r}\downarrow}
}{2}
\right\rangle_\tau
=\frac{n_x}{2}-\sigma m_x
\label{3.11}
\end{equation}
where $x$ again denotes the set (${\bf r}\tau$) while
$n_x$ and $-m_x$ are the expectation value of
the local number operator and the $z$-component of
the local spin operator  respectively.

The effective action or a generalization of the free
energy to the case of $\tau$-dependent dynamical variables
is defined by
\begin{equation}
F=\Omega+
\int^\beta_0 d\tau
\sum_{{\bf r}\sigma}
J_\sigma({\bf r}\tau)\phi_\sigma({\bf r}\tau)
\equiv
\Omega+\sum_{x\sigma} J_{x\sigma}\phi_{x\sigma}
\label{3.12}
\end{equation}
with an identity
\begin{equation}
J_{x\sigma}=\frac{\delta F}{\delta \phi_{x\sigma}}.
\label{3.13}
\end{equation}
$F$ corresponds to $\Gamma$ of the previous section.
The rule for $\phi$ corresponding to the rule (\ref{PHI2})
in this case is
\begin{equation}
\label{IPHI}
-\phi_\sigma
=
\left\langle
z_{\sigma}z^*_{\sigma}
e^{
-U
z^*_{\uparrow}z_{\uparrow}
z^*_{\downarrow}z_{\downarrow}
+\sum_{\sigma'}
(
J_{\sigma'}^{(1)}+
J_{\sigma'}^{(2)}+\cdots
)
z^*_{\sigma'}z_{\sigma'}
}
\right\rangle_{G^{(0)}},
\end{equation}
that is, the sum of all the connected graphs built with
4-point vertices $U$, pseudo-vertices $J^{(i)}_\sigma$
($i\geq 1$), and propagators $G^{(0)}_\sigma$
with the notation similar to (\ref{O}).
Here $G^{(0)}$ is defined as
\begin{equation}
\left[G^{(0)}_\sigma\right]^{-1}_{xy}
=
G^{-1}_{xy}-\delta_{xy}J^{(0)}_{x\sigma}.
\end{equation}
The extra minus sign in (\ref{IPHI}) originates from the sign
in $\phi=-\frac{\delta \Omega}{\delta J}$.
Then as mentioned before (below Proposition A$1$))
the inversion formula of the $n$-th order in $U$ is given
by the $n$-th order of (\ref{IPHI})
regarding both $\phi_\sigma$ and $G^{(0)}_\sigma$
as order unity.
Thus we obtain
\vspace{3pt}
\begin{equation}
\label{3.25}
\vspace{35pt}
\includegraphics{3-1.ps}
\end{equation}
\vspace{1pt}
\begin{equation}
\vspace{35pt}
\includegraphics{3-2.ps}
\label{3.26}
\end{equation}
\vspace{3pt}
and so on.
Here a solid (dashed) line to which an arrow is attached
(per loop of lines) represents the
propagator of spin-up (spin-down) electron
and it is $G^{(0)}_\uparrow$ ($G^{(0)}_\downarrow$).
The black dot denote the place where two propagators meet
(corresponding to a derivative with respect to $J^{(0)}$
-- note $\frac{\delta G^{(0)}_\sigma}{\delta J^{(0)}_{\sigma\prime}}
=\delta_{\sigma\sigma'}G^{(0)}_\sigma G^{(0)}_\sigma$).
The factor $U$ is associated with a 4-point vertex at which
spin-up and spin-down propagators come in and out, while
no factor is associated with the black dot (see Appendix B).
Hence from (\ref{3.26}) we get
\vspace{3pt}
\begin{equation}
\vspace{30pt}
\includegraphics{3-3.ps}
\end{equation}
or
\begin{equation}
J^{(1)}_{-\sigma}=U\phi_\sigma=-U{\rm Tr} G_\sigma.
\label{3.27}
\end{equation}
The second order formula of the inversion method is
also obtained as that order of (\ref{IPHI});
\vspace{20pt}
\\
\vspace{60pt}
\includegraphics{3-4.ps}
\begin{equation}
\label{3.28}
\end{equation}
\vspace{3pt}
which reduces to, as eq.~(\ref{23}) do to (\ref{J2}),
\vspace{3pt}
\begin{equation}
\vspace{30pt}
\includegraphics{3-5.ps}
\label{3.29}
\end{equation}
\vspace{3pt}
Further,
it is easy to find that, corresponding to (\ref{J3}),
\vspace{3pt}
\begin{equation}
\vspace{30pt}
\includegraphics{3-6.ps}
\label{3.30}
\end{equation}
\vspace{3pt}
The left-hand side of (\ref{3.29}) or (\ref{3.30})
can be written as
$\frac{\delta \phi_\uparrow}{\delta J^{(0)}_\uparrow}J^{(i)}_\uparrow$
with $i=2$ or 3.
Following the procedure presented in the previous section we get
\begin{eqnarray}
F &=&
F^{(0)}+
F^{(1)}+
F^{(2)}+
F^{(3)}+\cdots ,
\label{3.31}
\\
F^{(0)} &=&
\sum_{x\sigma}J^{(0)}_{x\sigma}\phi_{x\sigma}
-\sum_\sigma
{\rm Tr} \ln \left[ G^{(0)}_{\sigma} \right]^{-1},
\label{3.32}
\\
F^{(1)} &=&
U\sum_x \phi_{x\uparrow} \phi_{x\downarrow}
=
\frac{U}{2}\sum_{x\sigma}
\phi_{x\sigma} \phi_{x \: -\sigma},
\label{3.33}
\end{eqnarray}
\begin{equation}
\vspace{30pt}
\includegraphics{3-7.ps}
\label{3.34}
\end{equation}
\vspace{1pt}
\begin{equation}
\vspace{30pt}
\includegraphics{3-8.ps}
\label{3.35}
\end{equation}
\vspace{3pt}
where  $F^{(n)}$ satisfies
\begin{equation}
J^{(n)}_\sigma =
\frac{\delta F^{(n)}}{\delta \phi_\sigma}.
\nonumber
\end{equation}
Note that $J^{(0)}$ contained in
$G^{(0)}$ is a functional of $\phi$
defined by the solution of (\ref{3.25})
or
\begin{equation}
\phi_{x\sigma} = - G^{(0)}_{xx\sigma}=
-\left(\frac{1}{G^{-1}-J^{(0)}_{\sigma}[\phi]}
\right)_{xx}.
\label{3.25'}
\end{equation}
The free energy of the Stoner theory is recreated by
$F^{(0)}+F^{(1)}$.
Now it is clear that all the propositions given in
Sec.~\ref{s2} hold  for the present model with minor
and self-evident modifications.
Here we repeat them for later convenience.
\begin{description}
\item[Proposition B1)]
The graphical rule for $\Delta F$ is
given by the following equation;
\begin{equation}
\Delta F=-
\left\langle
e^{
-U
z^*_{\uparrow}z_{\uparrow}
z^*_{\downarrow}z_{\downarrow}
+\sum_{\sigma}
(J^{(2)}_{\sigma}
+J^{(3)}_{\sigma}+\cdots )
z^*_{\sigma}z_{\sigma}
}
\right\rangle_{G^{(0)}}^{\rm 1VI/nself.},
\end{equation}
that is, the sum of all the connected 1VI/nself.\ diagram
constructed out of 4-point vertices, 2-point pseudo-vertices
and propagators $G^{(0)}_\sigma$.
\end{description}
Here 1VI/nself.~condition implies that
only the 1VI graphs are kept and graphs corresponding
to the self-contractions of the vertices are excluded.
\begin{description}
\item[Proposition B$\bf 2$)]
$J^{(n)}$ is successively given as a functional of
$J^{(0)}_\sigma$ by the following formula.
\begin{equation}
J^{(n)}_\sigma
=
\mbox{ $n$-th order of}
\quad
D_\sigma \times
\left\langle
z^*_{\sigma}z_{\sigma}
e^{
-U
z^*_{\uparrow}z_{\uparrow}
z^*_{\downarrow}z_{\downarrow}
+\sum_{\sigma'}
(J^{(2)}_{\sigma'}+
J^{(3)}_{\sigma'}+\cdots )
z^*_{\sigma'}z_{\sigma'}
}
\right\rangle_{G^{(0)}}^{\rm 1VI/ndself.}.
\end{equation}
where
\begin{equation}
D^{-1}_\sigma=\frac{\delta \phi_\sigma}{\delta J^{(0)}_\sigma}
=
-\frac{1}{G^{-1}-J^{(0)}_\sigma}
\frac{1}{G^{-1}-J^{(0)}_\sigma}.
\end{equation}
\end{description}
\begin{description}
\item[Proposition B3)]
The graphical rule for $\Delta F$ is given by
the following formula
\begin{equation}
\Delta F=-
\left\langle
e^{
-U
z^*_{\uparrow}z_{\uparrow}
z^*_{\downarrow}z_{\downarrow}
+\sum_{\sigma}
z^*_{\sigma}z_{\sigma}
\varphi_{-\sigma}
}
\right\rangle_{G^{(0)},D}^{\rm tree/1VI/excl.},
\end{equation}
or, in a more detailed expression,
\begin{eqnarray}
\Delta F
&=&
-
\left.
\frac{\int {\cal D}z^*{\cal D}z{\cal D}\varphi e^{S_0+S_{int.}} }{
\int {\cal D}z^*{\cal D}z{\cal D}\varphi e^{S_0}}
\right|_{\rm conn./tree/1VI/excl.}
\nonumber \\
S_0
&=&
-\sum_\sigma z^*_\sigma [G^{(0)}]^{-1} z_\sigma
+\frac{1}{2}\sum_\sigma \varphi_\sigma D_{\sigma}^{-1} \varphi_\sigma
\nonumber \\
S_{int.}
&=&
-U
z^*_\uparrow
z_\uparrow
z^*_\downarrow
z_\downarrow
+\sum_\sigma
z^*_\sigma z_\sigma \varphi_{-\sigma}
\end{eqnarray}
where the subscript conn./tree/1VI/excl.\
implies that we should take only connected
graphs which are tree with respect to $D_\sigma$ propagator
of the bosonic field $\varphi_\sigma$ and
also 1VI with respect to the 4-point vertex and
the sub-structures of the graphs corresponding to
$\frac{\delta}{\delta J^{(0)}}{\rm Tr}\ln G^{(0)}$
and
$\frac{\delta^2}{\delta J^{(0)}\delta J^{(0)}}{\rm Tr}\ln G^{(0)}$
are excluded.
\end{description}
Note that Proposition B1) can be deduced from
the formula
\begin{equation}
\Delta F=-
\left\langle
e^{
-U
z^*_{\uparrow}z_{\uparrow}
z^*_{\downarrow}z_{\downarrow}
+\sum_{\sigma}
(J^{(2)}_{\sigma}
+J^{(3)}_{\sigma}+\cdots )
z^*_{\sigma}z_{\sigma}
}
\right\rangle_{G^{(0)}}^{\rm nself.},
\end{equation}
which is clear from the functional representation;
\begin{eqnarray}
e^{-F} &=&
e^{\sum_\sigma (
-J^{(0)}_\sigma\phi_\sigma+{\rm Tr} \ln [G^{(0)}_\sigma]^{-1} )
-U\phi_\uparrow \phi_\downarrow}
\nonumber
\\
&\times&
\frac{
\int {\cal D}z^* {\cal D}z
e^{
-\sum_\sigma z^*_\sigma [G^{(0)}_\sigma]^{-1} z_\sigma
-U
z^*_\uparrow z_\uparrow
z^*_\downarrow z_\downarrow
+\sum_\sigma [ (
J_\sigma -J^{(0)}_\sigma )
z^*_\sigma z_\sigma
-( J_\sigma -J^{(0)}_\sigma )
\phi_\sigma ]
+U\phi_\uparrow \phi_\downarrow
}
}{
\int {\cal D}z^* {\cal D}z
e^{
-\sum_\sigma z^*_\sigma [G^{(0)}_\sigma]^{-1} z_\sigma
}
}
\label{3.44}
\end{eqnarray}
or
\begin{equation}
e^{-\Delta F}=
\frac{
\int {\cal D}z^* {\cal D}z
e^{
-\sum_\sigma z^*_\sigma [G^{(0)}_\sigma]^{-1} z_\sigma
+( -U
z^*_\uparrow z_\uparrow
z^*_\downarrow z_\downarrow
+U\sum_\sigma \phi_{-\sigma}
z^*_\sigma z_\sigma
-U\phi_\uparrow \phi_\downarrow
)
-\sum_\sigma
\frac{\delta \Delta F}{\delta \phi_\sigma}
( z^*_\sigma z_\sigma -\phi_\sigma )
}
}{
\int {\cal D}z^* {\cal D}z
e^{
-\sum_\sigma z^*_\sigma [G^{(0)}]^{-1}_\sigma z_\sigma
}
}.
\label{3.45}
\end{equation}
There is another way to state the graph rule.
For this purpose
${\cal K}[A]$ is defined as follows:
\begin{equation}
{\cal K} [A]
=
\left\langle
e^{
-U
z^*_{\uparrow}z_{\uparrow}
z^*_{\downarrow}z_{\downarrow}
}
\right\rangle_A^{\rm 1VI}
\end{equation}
where $A$ is the propagator used in the graphical expression.
Then the rule is summarized in the following proposition.
\begin{description}
\item[Proposition B$\bf 1''$)]
$\Delta F =F-(F^{(0)}+F^{(1)})$ is given by
${\cal K}[\bar{G}]-\widetilde{\Delta F}$ where
\begin{equation}
\bar{G_\sigma}=
\left(
G^{-1}
-J^{(0)}_\sigma
-J^{(2)}_\sigma
-J^{(3)}_\sigma
-\cdots
\right)^{-1}
=
\left(
G^{-1} -J_\sigma +U\phi_{-\sigma}
\right)^{-1}
\label{3.36}
\end{equation}
and
\begin{eqnarray}
\widetilde{\Delta F}
&=&
\sum_\sigma {\rm Tr} \ln
( G^{-1}
 -J^{(0)}_\sigma
-J^{(2)}_\sigma
-J^{(3)}_\sigma
-\cdots )
\nonumber \\
&&
-\sum_\sigma {\rm Tr} \ln
( G^{-1}
-J^{(0)}_\sigma)
-\sum_\sigma \phi_\sigma
\left(J_\sigma -J^{(0)}_\sigma -J^{(1)}_\sigma\right).
\end{eqnarray}
In other words
\begin{eqnarray}
F &=&
\sum_\sigma \phi_\sigma J_\sigma
-U\phi_\uparrow\phi_\downarrow
-\sum_\sigma {\rm Tr} \ln \bar{G}^{-1}_\sigma
+{\cal K}[\bar{G}]
\nonumber \\
&\equiv&
\sum_\sigma \phi_\sigma J_\sigma
-U\phi_\uparrow\phi_\downarrow
+\bar{\cal K}[\bar{G}].
\end{eqnarray}
\end{description}
\setcounter{equation}{0}
\section{Case of QED}
\label{s4}
Final example is the effective action for the expectation value
of gauge invariant local composite field
$\phi^\mu(x)=\langle \bar{\psi}(x)\gamma^\mu\psi(x) \rangle$
in QED.
The practical use of $\Gamma[\phi_\mu]$ in QED is
as follows.
Although
$\langle \bar{\psi}\gamma^\mu\psi \rangle =0$
for the vacuum, the lowest relation of the
 on-shell condition\cite{ONSHELL} (with
 the space-time integration over $y$ and the
 summation over $\nu$ suppressed),
\begin{equation}
\Gamma^{(2)}_{\mu x, \nu y}
\Delta \phi^\nu(y)=0,
\end{equation}
where
\begin{equation}
\Gamma^{(2)}_{\mu x, \nu y}\equiv
\left.
\frac{\delta^2 \Gamma[\phi]}{
\delta \phi_\mu(x)
\delta \phi_\nu(y)}
\right|_{\phi=0},
\end{equation}
determines the bound state in the channel specified
by $\bar{\psi}\gamma^\mu\psi$.
This allows us a {\em gauge invariant study of} $^3S_1$
of positronium state.
The following work may also be a starting point for the
study of the order parameter for the chiral
symmetry breaking
$\phi=\langle \bar{\psi}\psi \rangle$
in the massless QED and that of
$\langle \bar{q}^a q^a \rangle$
or
$\langle A^a_\mu A^a_\mu \rangle$ in QCD.
Here $q$ and $A^a_\mu$ are operators for quarks and gluons
respectively.
All these are believed to be non-vanishing objects in
contrast to
$\bar{\psi}\gamma^\mu \psi$.
The lowest order discussion of $\langle \bar{\psi}\psi \rangle$
has been given in \cite{MLQED}.

The generating functional in this case is given by (with
the space-time integration and the summation over the Greek
index suppressed),
\begin{equation}
e^{iW[J,K]}=\int
{\cal D}\bar{\psi}
{\cal D}\psi
{\cal D}A
e^{iS[\bar{\psi}\psi A J]},
\label{4.1}
\end{equation}
\begin{eqnarray}
S[\bar{\psi},\psi, A, J]
&=&
-\bar{\psi}G^{-1} \psi
-\frac{1}{2}A^\mu D^{-1}_{\mu\nu}A^\nu
+e\bar{\psi}\gamma_\mu \psi A^\mu
+J_\mu \bar{\psi}\gamma^\mu \psi
\nonumber \\
&=&
-\bar{\psi}G^{-1}_J\psi
-\frac{1}{2}A^\mu D^{-1}_{\mu\nu}A^\nu
+e j_\mu  A^\mu
\label{4.2}
\end{eqnarray}
where
\begin{eqnarray}
G^{-1}
&=&
-i\gamma_\mu \partial^\mu+m,
\\
G_J^{-1}
&=&
G^{-1}-J_\mu \gamma^\mu,
\label{4.3} \\
D^{-1}_{\mu\nu}&=&
-\Box g_{\mu\nu}+\left(
1-\frac{1}{\lambda}\right)\partial_\mu\partial_\nu,
\label{4.4} \\
j_\mu&=&\bar{\psi}\gamma_\mu\psi.
\end{eqnarray}
Here the parameter $\lambda$ specifies the gauge.
Then we get Feynman graphs for
$\phi_\mu=\frac{\delta W}{\delta J^\mu}=\langle j_\mu \rangle
=\langle \bar{\psi}\gamma_\mu \psi \rangle$:
\begin{equation}
\label{PM}
\phi_\mu=
\left\langle
\bar{\psi}\gamma_\mu \psi
e^{ e
\bar{\psi}\gamma_\mu \psi
A^\mu
+
( J^{(1)}_\mu+
J^{(2)}_\mu+\cdots)
\bar{\psi}\gamma^\mu \psi
}
\right\rangle_{G^{(0)},D},
\end{equation}
that is, the sum of all the connected graphs built with 3-point
vertices
($\bar{\psi}\gamma^\mu\psi A_\mu$),
2-point pseudo-vertices
($J^{(i)}_\mu \bar{\psi}\gamma^\mu\psi$),
electron propagators $G^{(0)}$ and photon propagators $D$.
Here $J^{(i)}_\mu$ is the $i$-th order (in $e^2$)
of $J^\mu$ and the propagator
$G^{(0)}$ is $G_J$ but $J$ replaced by $J^{(0)}$.
The quantity $J^{(0)}$ is defined by
\begin{equation}
\phi_\mu(x)=i\gamma_\mu G^{(0)}_{\mu\mu}(x,x),
\end{equation}
which is equivalent to (\ref{P0}) below.
By writing down the $i$-th order of
(\ref{PM}) one gets the inversion formula
of that order.
For example,
\vspace{5pt}
\begin{equation}
\includegraphics{4-1.ps}
\label{P0}
\end{equation}
\vspace{5pt}
\begin{equation}
\includegraphics{4-2.ps}
\label{P1}
\end{equation}
\vspace{15pt}
\begin{equation}
\vspace{30pt}
\includegraphics{4-3.ps}
\label{P2}
\end{equation}
\vspace{190pt}
In the above graphs we associate
the  electron propagator $G^{(0)}$
with each solid line and the photon propagator $D$
with each dashed line.
In addition a factor $e\gamma_{\mu}$ and $\gamma^\mu$ are assigned
to a vertex  and to a black dot ($\bullet$) respectively
(see Appendix B).

If we define $J^{(1)}_A$ and $J^{(1)}_B$ from (\ref{P1}) by
\begin{equation}
\vspace{30pt}
\includegraphics{4-4.ps}
\label{J1Q}
\end{equation}
we see that all the
$J^{(1)}_A$'s exactly cancel out the
1PR structure appearing in the $i$-th
order of (\ref{PM}) with $i\geq 2$.
Here 1PR means 1-particle-reducible
{\em with respect to the photon propagator}.
Hereafter  1PI graph is defined as the
graph which is not 1PR {\em in photon channel}.
Indeed  all the 1PR graph in (\ref{P2})
disappear after substitution of the last equation
due to $J^{(1)}_A$ while $J^{(1)}_B$ remains;
\begin{equation}
\vspace{30pt}
\includegraphics{4-5.ps}
\label{J2Q}
\end{equation}
\vspace{70pt}

The effective action in this case is defined by
$\Gamma=W-J_\mu\phi^\mu$
(with the space-time integration suppressed) as usual with an identity
$-J^\mu=\frac{\delta \Gamma}{\delta \phi_\mu}$.
Thus integrating (\ref{J2Q})
one can obtain $\Gamma^{(2)}$ (and higher order
by using (\ref{PM})).
Here we can take another course instead.
For this purpose let us first examine
the path-integral representation of $\Gamma$.
Integrating out the photon field we get
\begin{equation}
e^{i\Gamma}=
\int {\cal D}\bar{\psi} {\cal D}\psi
e^{-i\bar{\psi}G^{-1} \psi
+i\frac{e^2}{2}j_\mu D^{\mu\nu} j_\nu
+iJ_\mu j^\mu
-iJ_\mu \phi^\mu}.
\end{equation}
Since $\Gamma^{(i)}$ is defined by
$-J^{(i)}_\mu = \frac{\delta \Gamma^{(i)}}{\delta \phi^\mu}$
the quantities $\Gamma^{(1)}_A$ and $\Gamma^{(1)}_B$ are defined
as
\begin{equation}
\vspace{30pt}
\includegraphics{4-6.ps}
\end{equation}
in accordance with (\ref{J1Q}).
The quantity
$\Delta J$ and $\Delta \Gamma$ in this case are expanded as
\begin{equation}
\Delta J = J^{(1)}_B+J^{(2)}+J^{(3)}+\cdots,
\end{equation}
\begin{equation}
\Delta \Gamma = \Gamma^{(1)}_B+\Gamma^{(2)}+\Gamma^{(3)}+\cdots.
\end{equation}
Noting that
$\Gamma^{(1)}_A=\frac{e^2}{2}\phi^\mu D_{\mu\nu} \phi^\nu$
and
$\Gamma^{(0)}=
-J^{(0)}_\mu \phi^\mu
-i {\rm Tr}\ln [G^{(0)}]^{-1}$, we get
\begin{eqnarray}
\lefteqn{e^{i\Gamma}
=
e^{i
( -J^{(0)}_\mu \phi^\mu
-i {\rm Tr}\ln [G^{(0)}]^{-1})
+i\frac{e^2}{2}\phi^\mu D_{\mu\nu}\phi^\nu}
}
\nonumber \\
&=&
\frac{
\int {\cal D}\bar{\psi}{\cal D}\psi
e^{
-i\bar{\psi}[G^{(0)}]^{-1}\psi
+i\frac{e^2}{2}j^\mu D_{\mu\nu}j^\nu
+i(J_\mu-J^{(0)}_\mu)j^\mu
-i(J_\mu-J^{(0)}_\mu
+\frac{e^2}{2}\phi^\nu D_{\nu\mu})\phi^\mu
}
}{
\int {\cal D}\bar{\psi}{\cal D}\psi
e^{
-i\bar{\psi}[G^{(0)}]^{-1}\psi
}
}
\end{eqnarray}
or
\begin{equation}
e^{i\Delta \Gamma}
=
\frac{
\int {\cal D}\bar{\psi}{\cal D}\psi
e^{
i[-\bar{\psi}[G^{(0)}]^{-1}\psi
+e^2({\frac{1}{2}j^\mu D_{\mu\nu}j^\nu
-\phi^\mu D_{\mu\nu} j^\mu)
+\frac{1}{2}\phi^\mu D_{\mu\nu}\phi^\nu}
-\frac{\delta \Delta \Gamma}{\delta \phi^\mu}
\left( j^\mu-\phi^\mu\right)]
}
}{
\int {\cal D}\bar{\psi}{\cal D}\psi
e^{
-i\bar{\psi}[G^{(0)}]^{-1}\psi
}
}.
\label{4.13}
\end{equation}
We write (\ref{4.13}) as
\begin{equation}
\Delta \Gamma[\phi]=
\frac{1}{i}
\left\langle
e^{i\frac{e^2}{2}j_\mu D^{\mu\nu} j_\nu-
i\frac{\delta \Delta \Gamma}{\delta \phi^\mu}j_\mu}
\right\rangle^{\rm nself.}_{G^{(0)}}.
\label{4.14}
\end{equation}
The meaning of nself.\
is that we have to exclude the self-contraction of
the electron propagators.
By using (\ref{4.14}) and noting
the cancellation similar to that in (\ref{P2})
we get
\begin{equation}
\vspace{30pt}
\includegraphics{4-7.ps}
\end{equation}
\vspace{60pt}
Combined with the arguments similar to those of
previous subsections, we arrive at the
following proposition (with a similar statement
for a graph of $J^{(n)}$):
\begin{description}
\item[Proposition C)] $\Gamma^{(n)}$ ($n\geq2$)
is the sum of all possible $n$-th order (in $e^2$)
1PI diagram constructed out of 4-point vertex of
order $e^2$
(
\setlength{\unitlength}{1 mm}
\begin{picture}(20,10)
\thicklines
\put(5,5){\vector(0,-1){5}}
\put(5,0){\line(0,-1){5}}
\multiput(5,0)(2,0){6}{\line(1,0){1}}
\put(16,0){\line(0,1){5}}
\put(16,-5){\vector(0,1){5}}
\end{picture}
), vertex of order $e^2$
(
\setlength{\unitlength}{1 mm}
\begin{picture}(25,10)
\thicklines
\put(0,0){\line(1,0){20}}
\put(10,0){\circle*{1}}
\put(7,0){\vector(1,0){0}}
\put(10,3){\makebox(0,0)[b]{$J^{(1)}_B$}}
\end{picture}
)
and vertices of order $e^{2i}$
(
\setlength{\unitlength}{1 mm}
\begin{picture}(25,10)
\thicklines
\put(0,0){\line(1,0){20}}
\put(10,0){\circle*{1}}
\put(7,0){\vector(1,0){0}}
\put(10,1){\makebox(0,0)[b]{$J^{(i)}$}}
\end{picture}
) ($2\leq i < n$).
Here the propagator is $G^{(0)}$.
In other words
\begin{equation}
\Delta \Gamma[\phi]=
\frac{1}{i} \left\langle
e^{\frac{ie^2}{2}j_\mu D^{\mu\nu} j_\mu
+i(
J^{(1)}_{B\mu}
+J^{(2)}_\mu
+J^{(3)}_\mu+\cdots)j^\mu
}\right\rangle_{G^{(0)}}^{\rm 1PI}
\end{equation}
where $\langle \cdots \rangle_{\rm 1PI}$
means 1PI (in terms of the photon lines)
connected Wick contraction using the propagators
$G^{(0)}$ that is a functional of $J^{(0)}$.
\end{description}
Of course, there are various equivalent modifications of
Proposition C).
\section*{Acknowledgment}

I would like to thank  Professor R. Fukuda and
Mr. S. Yokojima
 for  suggestions, discussions, and encouragement
throughout the work.
\setcounter{equation}{0}
\renewcommand{\theequation}{A.\arabic{equation}}
\section*{Appendix A --- Legendre transformation and the inversion method}
In this appendix we look at more carefully the reason why we should
assume that $\phi$ is of order $\lambda^0=1$ or independent of $\lambda$
in our inversion process. This point has been exemplified in terms
of diagrams, which is not necessarily familiar to everyone.
Here we present a clear explanation in purely mathematical
language.
Although the following discussion is trivial it is
worth while in order to understand the foundation of the
inversion method.
For brevity the case of $x$-independent variables $J$ and
$\phi$ are considered.

Consider the quantity
$W[J,\lambda]-J\phi[J,\lambda]$
in which
$\phi[J,\lambda]\equiv\frac{\delta W[J,\lambda]}{\delta J}$.
Here we have emphasized the $\lambda$-dependence. If we take a small
variation of this quantity {\em assuming that
$J$ and $\lambda$ are independent variables} it becomes
\begin{equation}
\frac{\delta W[J,\lambda]}{\delta J}dJ
+
\frac{\delta W[J,\lambda]}{\delta \lambda}d\lambda
-
dJ\phi[J,\lambda]-Jd\phi[J,\lambda]
=
\frac{\delta W[J,\lambda]}{\delta \lambda}d\lambda
-Jd\phi[J,\lambda].
\label{ap1}
\end{equation}
Hence we see that the quantity can be regarded as a function(al) of
two independent variables $\phi$ and $\lambda$. We thus
write the quantity $W-J\phi$ as
$\Gamma[\phi,\lambda]$. What is implied here is as follows: if
we solve the relation $\phi=\frac{\delta W[J,\lambda]}{\delta J}$
in favor of $J$ {\em assuming that the two quantities $\phi$
and $\lambda$ are mutually independent} to get $J=J[\phi,\lambda]$
and then insert this expression of $J$ into all $J$ appearing in
$W[J]-J\phi$, then $W[J]-J\phi$ is automatically written by only
two independent variables $\phi$ and $\lambda$.
In other words, the inversion process of Legendre transformation
is carried out regarding $\phi$ as independent of $\lambda$.
Hence the process in the inversion method exactly coincides with
the inversion process of Legendre transformation.
Note that once the inversion or Legendre transformation is performed
and after the sources are set to the desired values, zero
for example, the resultant
$\phi$ depends on $\lambda$ of course.
\setcounter{equation}{0}
\renewcommand{\theequation}{B.\arabic{equation}}
\section*{Appendix B --- Feynman rules}
\subsubsection*{$\varphi^4$ theory}
Although well known, we summarize for clarity the
rule (Rule A) to get algebraic
expressions from the corresponding
graphs for the $\varphi^4$ theory.
\begin{description}
\item[Rule A1)] In one
specific way (as one likes),
assign $n$ labels $x_{1},\ldots,x_{n}$ (internal points)
to all the 4-point vertices and the pseudo-vertices
where $n$ is the total number of vertices (including
pseudo-vertex).
\item[Rule A2)]
Associate a propagator
$G_J$ (for the rule (\ref{W1}) and (\ref{PHI1}))
or
$G^{(0)}$  (for the rule (\ref{W2}) and (\ref{PHI2}))
with each line.
A factor $-\lambda$, and $J^{(i)}$
are assigned to the 4-point
vertex and the pseudo-vertex of the $i$-th order respectively.
No factor is assigned to the black dot which corresponds
external point.
\item[Rule A3)]
Associate a factor $i^{-L}$ for a diagram
where $L$ is the number of independent momenta of the graph.
\item[Rule A4)]
Associate a symmetry factor $S$ for a diagram.
\item[Rule A5)]
Sum (Integrate) the product of all factors
in {\bf A2)} to {\bf A4)} over the space time index
$x_{1},\ldots,x_{n}$.
\end{description}
The symmetry factor $S$ for each graph is given by the
line symmetry number $S_{L}$ and the vertex symmetry
number $S_{V}$ as
$S=
\frac{1}{S_{L}}\cdot
\frac{1}{S_{V}}$.
As is well known, $S_{L}$ and $S_{V}$ are obtained
through the following rule.
\begin{description}
\item[Rule $\bf S_{L}$1)]
If there is a line which starts from a
vertex (including the black dot $\bullet$ and pseudo-vertex) and come
back directly to the starting vertex, associate the factor $2$.
\item[Rule $\bf S_{L}$2)]
If there are $m$ lines ($m=2,3,4$) directly connecting
two common vertices (including pseudo-vertex),
associate the factor $m!$.
\item[Rule $\bf S_{L}$3)]
All the product of the factors in
$\bf S_{L}1)$ and $\bf S_{L}2)$ is $S_{L}$.
\item[Rule $\bf S_{V}$)]
Assign $n$ labels $1,\ldots,n$ to $n$ vertices (including
pseudo-vertex)
in an arbitrary way.
Count the number of all possible other ways of assigning
$n$ labels that give the same topological structure
as the first specific way. The number thus obtained
plus $1$ is $S_{V}$.
\end{description}
For definiteness we give some examples;
the graph appearing in (\ref{10}) has
($S_{L},S_{V}$)=($2,1$):
three graphs of (\ref{12}) have
($2!^{2}\cdot 2,1$), ($2^{2}, 2$), and ($3!, 2$)
respectively.

As another example we consider the
reduction of (\ref{23}) to (\ref{J2}).
Since the symmetry factors of the second, fourth, and sixth
graphs on the left-hand side of (\ref{23}) are
($S_{L},S_{V}$)=($1,2$), ($2,1$), and ($2^2,2$),
the contribution of the three graphs becomes zero.
This is
because, after replacing $J^{(1)}$ by use of (\ref{21})
(whose symmetry factor is $2$),
new symmetry factors of these graphs becomes
$1\cdot 2 \cdot 2^2$, $2\cdot 1 \cdot 2$, and $2^2\cdot 2$
respectively.
By a similar argument we find the cancellation of the
third and fifth graphs on the left-hand side of (\ref{23}).
Thus we get (\ref{J2}) from (\ref{23}).


\subsubsection*{Itinerant electron model}
The rules for the itinerant electron model
are given as follows (Rule B).
Rule B1), B4) and B5) are the same as
Rule A1), A4) and A5) respectively.
Rule B3) is Rule A3) with $i^{-L}$ replaced by
$(-1)^L(-1)^{L_f}$ where $L_f$ is the number of
Fermion loops. Rule A2) is changed into
\begin{description}
\item[Rule B2)]
Associate
\setlength{\unitlength}{1 mm}
\begin{picture}(20,5)
\thicklines
\put(11,0){\vector(1,0){0}}
\put(5,0){\line(1,0){10}}
\put(5,0){\makebox(0,0)[r]{$y$}}
\put(15,0){\makebox(0,0)[l]{$x$}}
\end{picture}
and
\begin{picture}(20,5)
\thicklines
\put(11,0){\vector(1,0){0}}
\multiput(5,0)(2,0){6}{\line(1,0){1}}
\put(5,0){\makebox(0,0)[r]{$y$}}
\put(16,0){\makebox(0,0)[l]{$x$}}
\end{picture}
with $[G^{(0)}_\uparrow]_{xy}$ and $[G^{(0)}_\downarrow]_{xy}$
respectively and  the factor $U$ is assigned to
the 4-point vertex.
The factor $J^{(i)}_\sigma$ is also associated with
the pseudo-vertex of the $i$-th order.
No factor is assigned to the external point.
\end{description}
As for the symmetry factor
$S$($=\frac{1}{S_L}\cdot \frac{1}{S_V}$)
rules for $S_L$ and $S_V$ is essentially the same
as those of the $\varphi^4$ theory
except for the fact that we have to distinguish the spin-up
and spin-down propagators and their directions
of the arrows when we consider the topological
equivalence.
Thus the factor $S_L$ is always 1 in this model.

\subsubsection*{QED}
Finally the rules for QED are presented as follows.
\begin{description}
\item[Rule C1)] Assign $n$
labels, in one specific way as one likes, ($x_1,\mu_1$), \ldots
($x_{n},\mu_{n}$) to its vertices (including pseudo-vertices).
\item[Rule C2)] Associate an electron propagator $G^{(0)}$
with each solid line and a photon propagator $D$
with each dashed line.
\item[Rule C3)] Associate a factor $e\gamma_{\mu}$ and
$J^{(i)}_\mu\gamma^\mu$
with the 3-point vertex and the pseudo-vertex
respectively.
$\gamma^{\mu}$  is assigned to
the black dot ($\bullet$) which corresponds the external
point.
\item[Rule C4)] Associate a factor $i^{-L}(-1)^{L_f}$
where $L$ is the number of loop momenta of the graph
and $L_f$ is the number of the fermion loops.
\item[Rule C5)] Associate a symmetry factor $S$ for a diagram.
\item[Rule C6)] Sum the product of all factors in {\bf C2)}
to {\bf C5)} over the $x_1\cdots x_{n}$ and
$\mu_1 \cdots \mu_{n}$.
\end{description}
The symmetry factors are calculated as
before.
Note that $S_L$ is always 1 in QED.
\setcounter{equation}{0}
\renewcommand{\theequation}{C.\arabic{equation}}
\section*{Appendix C --- Inversion method for
$\langle \varphi(x)\rangle$ and
$\langle \varphi(x)\varphi(y)\rangle$ }
We show below how the inversion method works to reproduce
well-known results for the effective action of
$\langle\varphi(x)\rangle$ and
$\langle\varphi(x)\varphi(y)\rangle$.
For simplicity we consider the $\varphi^4$ theory
and several lower orders of the known rule are explicitly
studied rather than giving formal proof.
\subsubsection*{Case of $\langle\varphi(x)\rangle$}
In order to study the effective action of elementary field
$\phi(x)$, the generating functional $W[J]$ is defined as
in (\ref{1}) with $S[\varphi,J]$ replaced by
\begin{equation}
S[\varphi,J] =
- \frac{1}{2} \int d^{4} x
\varphi(x) (\Box+m^{2}) \varphi(x)
- \frac{\lambda}{4!} \int d^{4} x \varphi(x)^{4}
+ \int d^{4} x
J(x) \varphi(x).
\end{equation}
The dynamical variable $\phi$  for the effective action is
\begin{equation}
\phi(x)=\frac{\delta W}{\delta J(x)}
\equiv \langle \varphi(x) \rangle^{J}
\end{equation}
by use of which $\Gamma[\phi]$ is defined by (\ref{7}) and
eq.~(\ref{8}) holds as an identity.

Now the original series expansion in $\lambda$
is given by, suppressing the $x$-dependence,
\begin{equation}
\phi^{(0)}=
\setlength{\unitlength}{1mm}
\begin{picture}(15,0)(-1,0)
\thicklines
\put(0,1){\line(1,0){10}}
\put(10,1){\circle*{1}}
\end{picture},
\label{a3}
\end{equation}
\begin{equation}
\phi^{(1)}=
\setlength{\unitlength}{1mm}
\begin{picture}(35,8)(-1,-1)
\thicklines
\put(0,0){\line(1,0){10}}
\put(10,0){\circle*{1}}
\put(5,2){\circle{4}}
\put(15,0){\makebox(0,0){+}}
\put(20,0){\line(1,0){10}}
\put(25,3){\line(0,-1){6}}
\put(25,3){\circle*{1}}
\put(25,-3){\circle*{1}}
\put(30,0){\circle*{1}}
\end{picture},
\label{a4}
\end{equation}
\begin{eqnarray}
\phi^{(2)}&=&
\setlength{\unitlength}{1mm}
\begin{picture}(15,8)(-1,-1)
\thicklines
\put(0,0){\line(1,0){15}}
\put(15,0){\circle*{1}}
\put(5,3){\line(0,-1){6}}
\put(5,3){\circle*{1}}
\put(5,-3){\circle*{1}}
\put(10,3){\line(0,-1){6}}
\put(10,3){\circle*{1}}
\put(10,-3){\circle*{1}}
\put(20,0){\makebox(0,0){+}}
\put(25,0){\line(1,0){15}}
\put(40,0){\circle*{1}}
\put(30,2){\circle{4}}
\put(35,3){\line(0,-1){6}}
\put(35,3){\circle*{1}}
\put(35,-3){\circle*{1}}
\put(45,0){\makebox(0,0){+}}
\put(50,0){\line(1,0){15}}
\put(65,0){\circle*{1}}
\put(55,3){\line(0,-1){6}}
\put(55,3){\circle*{1}}
\put(55,-3){\circle*{1}}
\put(60,2){\circle{4}}
\put(70,0){\makebox(0,0){+}}
\put(75,0){\line(1,0){15}}
\put(90,0){\circle*{1}}
\put(80,2){\circle{4}}
\put(85,2){\circle{4}}
\end{picture}
\nonumber \\
&+&
\setlength{\unitlength}{1mm}
\begin{picture}(65,15)(-1,-1)
\thicklines
\put(0,0){\line(1,0){15}}
\put(15,0){\circle*{1}}
\put(7.5,0){\oval(4,4)[t]}
\put(5.5,0){\line(0,-1){3}}
\put(9.5,0){\line(0,-1){3}}
\put(5.5,-3){\circle*{1}}
\put(9.5,-3){\circle*{1}}
\put(20,0){\makebox(0,0){+}}
\put(25,0){\line(1,0){10}}
\put(35,0){\circle*{1}}
\put(30,2){\circle{4}}
\put(30,6){\circle{4}}
\put(40,0){\makebox(0,0){+}}
\put(45,0){\line(1,0){15}}
\put(60,0){\circle*{1}}
\put(52.5,0){\circle{6}}
\end{picture}.
\label{a5}
\end{eqnarray}
Here a black dot denotes the external source $J$
and a line the propagator $\frac{1}{\Box+m^2}$.
Thus (\ref{16}), the right-hand side of which is (\ref{a3}) with
$J$ replaced by $J^{(0)}$, becomes
\begin{equation}
\phi=
\setlength{\unitlength}{1mm}
\begin{picture}(15,8)(-1,-1)
\thicklines
\put(0,0){\line(1,0){10}}
\put(10,0){\circle*{1}}
\end{picture}
\label{a6}
\end{equation}
or
\begin{equation}
\phi(x)=\left(\frac{1}{\Box+m^2}\right)_{xy}J^{(0)}(y)
\label{a7}
\end{equation}
from which $J^{(0)}$ is obtained explicitly as
opposed to the case of the local composite operators;
\begin{equation}
J^{(0)}=\left(\Box+m^2\right)_{xy}\phi(y).
\end{equation}
Hereafter, all the black dots in the
graphs denote $J^{(0)}$ instead of $J$ as in (\ref{a6}).
We immediately know
\begin{equation}
\Gamma^{(0)}=-\frac{1}{2}\phi\left(\Box+m^2\right)\phi
\label{a9}
\end{equation}
by integrating $J^{(0)}=-\frac{\delta \Gamma^{(0)}}{\delta \phi}$.
 From (\ref{a3}) and (\ref{a4}),
the inversion formula of order $\lambda$, (\ref{17}), becomes
\begin{equation}
\setlength{\unitlength}{1mm}
\begin{picture}(15,8)(-1,-1)
\thicklines
\put(0,0){\line(1,0){10}}
\put(13,0){\makebox(0,0){$J^{(1)}$}}
\put(20,0){\makebox(0,0){+}}
\put(25,0){\line(1,0){10}}
\put(35,0){\circle*{1}}
\put(30,3){\line(0,-1){6}}
\put(30,3){\circle*{1}}
\put(30,-3){\circle*{1}}
\put(40,0){\makebox(0,0){+}}
\put(45,0){\line(1,0){10}}
\put(55,0){\circle*{1}}
\put(50,2){\circle{4}}
\put(60,0){\makebox(0,0){=0}}
\end{picture}
\label{a10}
\end{equation}
by noting that $\phi^{(0)\prime}=\left(\Box+m^2\right)^{-1}$,
which is denoted by a line.
The integration of $J^{(1)}=-\frac{\delta \Gamma^{(1)}}{\delta \phi}$
leads to
\begin{equation}
\Gamma^{(1)}=
\setlength{\unitlength}{1mm}
\begin{picture}(35,13)(-1,-1)
\thicklines
\put(0,0){\line(1,0){10}}
\put(0,0){\circle*{1}}
\put(10,0){\circle*{1}}
\put(5,3){\line(0,-1){6}}
\put(5,3){\circle*{1}}
\put(5,-3){\circle*{1}}
\put(15,0){\makebox(0,0){+}}
\put(20,0){\line(1,0){10}}
\put(20,0){\circle*{1}}
\put(30,0){\circle*{1}}
\put(25,2){\circle{4}}
\end{picture}.
\label{a11}
\end{equation}
By (\ref{a6}) we confirm that $\Gamma^{(1)}$ is a functional of
$\phi$ indeed.
Equation (\ref{a9}) and the first term in (\ref{a11})
constitute the usual tree part of the 1PI (1-particle-irreducible)
effective action.
 From (\ref{a3}) to (\ref{a5}),
the second order formula (\ref{18}) is written as
\begin{eqnarray}
\lefteqn{
\setlength{\unitlength}{1mm}
\begin{picture}(15,8)(-1,-1)
\thicklines
\put(0,0){\line(1,0){10}}
\put(13,0){\makebox(0,0){$J^{(2)}$}}
\put(20,0){\makebox(0,0){+}}
\put(25,0){\line(1,0){10}}
\put(38,0){\makebox(0,0){$J^{(1)}$}}
\put(30,2){\circle{4}}
\put(45,0){\makebox(0,0){+}}
\put(50,0){\line(1,0){10}}
\put(63,0){\makebox(0,0){$J^{(1)}$}}
\put(55,3){\line(0,-1){6}}
\put(55,3){\circle*{1}}
\put(55,-3){\circle*{1}}
\end{picture}
} \nonumber \\
&+&
\setlength{\unitlength}{1mm}
\begin{picture}(15,15)(-1,-1)
\thicklines
\put(0,0){\line(1,0){15}}
\put(15,0){\circle*{1}}
\put(5,3){\line(0,-1){6}}
\put(5,3){\circle*{1}}
\put(5,-3){\circle*{1}}
\put(10,3){\line(0,-1){6}}
\put(10,3){\circle*{1}}
\put(10,-3){\circle*{1}}
\put(20,0){\makebox(0,0){+}}
\put(25,0){\line(1,0){15}}
\put(40,0){\circle*{1}}
\put(30,2){\circle{4}}
\put(35,3){\line(0,-1){6}}
\put(35,3){\circle*{1}}
\put(35,-3){\circle*{1}}
\put(45,0){\makebox(0,0){+}}
\put(50,0){\line(1,0){15}}
\put(65,0){\circle*{1}}
\put(55,3){\line(0,-1){6}}
\put(55,3){\circle*{1}}
\put(55,-3){\circle*{1}}
\put(60,2){\circle{4}}
\put(70,0){\makebox(0,0){+}}
\put(75,0){\line(1,0){15}}
\put(90,0){\circle*{1}}
\put(80,2){\circle{4}}
\put(85,2){\circle{4}}
\end{picture}
\nonumber \\
&+&
\setlength{\unitlength}{1mm}
\begin{picture}(15,15)(-1,-1)
\thicklines
\put(0,0){\line(1,0){15}}
\put(15,0){\circle*{1}}
\put(7.5,0){\oval(4,4)[t]}
\put(5.5,0){\line(0,-1){3}}
\put(9.5,0){\line(0,-1){3}}
\put(5.5,-3){\circle*{1}}
\put(9.5,-3){\circle*{1}}
\put(20,0){\makebox(0,0){+}}
\put(25,0){\line(1,0){10}}
\put(35,0){\circle*{1}}
\put(30,2){\circle{4}}
\put(30,6){\circle{4}}
\put(40,0){\makebox(0,0){+}}
\put(45,0){\line(1,0){15}}
\put(60,0){\circle*{1}}
\put(52.5,0){\circle{6}}
\put(65,0){\makebox(0,0){$ = 0 .$}}
\end{picture}
\label{a12}
\end{eqnarray}
The second term in (\ref{18}) disappears because
$\phi^{(0)\prime\prime}[J^{(0)}]=0$.
Using (\ref{a10}) we see that 1-particle-reducible (1PR)
graphs in (\ref{a12}) exactly cancel out each other to get
\begin{equation}
\setlength{\unitlength}{1mm}
\begin{picture}(15,8)(-1,-1)
\thicklines
\put(0,0){\line(1,0){10}}
\put(13,0){\makebox(0,0){$J^{(2)}$}}
\put(20,0){\makebox(0,0){+}}
\put(25,0){\line(1,0){15}}
\put(40,0){\circle*{1}}
\put(32.5,0){\oval(4,4)[t]}
\put(30.5,0){\line(0,-1){3}}
\put(34.5,0){\line(0,-1){3}}
\put(30.5,-3){\circle*{1}}
\put(34.5,-3){\circle*{1}}
\put(45,0){\makebox(0,0){+}}
\put(50,0){\line(1,0){10}}
\put(60,0){\circle*{1}}
\put(55,2){\circle{4}}
\put(55,6){\circle{4}}
\put(65,0){\makebox(0,0){+}}
\put(70,0){\line(1,0){15}}
\put(85,0){\circle*{1}}
\put(77.5,0){\circle{6}}
\put(93,0){\makebox(0,0){$=0$}}
\end{picture}
\label{a13}
\end{equation}
from which we obtain
\begin{equation}
\Gamma^{(2)}=
\setlength{\unitlength}{1mm}
\begin{picture}(60,8)(-1,-1)
\thicklines
\put(0,2){\line(1,0){10}}
\put(0,-2){\line(1,0){10}}
\put(0,2){\circle*{1}}
\put(0,-2){\circle*{1}}
\put(10,2){\circle*{1}}
\put(10,-2){\circle*{1}}
\put(5,0){\circle{4}}
\put(15,0){\makebox(0,0){+}}
\put(20,0){\line(1,0){10}}
\put(30,0){\circle*{1}}
\put(20,0){\circle*{1}}
\put(25,2){\circle{4}}
\put(25,6){\circle{4}}
\put(35,0){\makebox(0,0){+}}
\put(40,0){\line(1,0){15}}
\put(40,0){\circle*{1}}
\put(55,0){\circle*{1}}
\put(47.5,0){\circle{6}}
\end{picture}.
\label{a14}
\end{equation}
This course of study can be continued up to  desired order
to give the well-known result;
\begin{equation}
\Gamma=-\frac{1}{2}\phi\left(\Box+m^2\right)\phi -
\frac{\lambda}{4!}\phi^4+{\cal K}_{\rm 1PI}[\phi]
\end{equation}
where ${\cal K}_{\rm 1PI}[\phi]$ is 1PI vacuum graph
${\cal K}_{\rm 1PI}[(\Box+m^2)^{-1}J]$
(written in terms of original $J$-representation)
but with $(\Box+m^2)^{-1}J$ replaced by $\phi$ or
\begin{equation}
{\cal K}_{\rm 1PI}[\phi]=
\setlength{\unitlength}{1mm}
\begin{picture}(15,8)(-1,-1)
\thicklines
\put(0,0){\line(1,0){10}}
\put(0,0){\circle*{1}}
\put(10,0){\circle*{1}}
\put(5,2){\circle{4}}
\put(15,0){\makebox(0,0){+}}
\put(20,2){\line(1,0){10}}
\put(20,-2){\line(1,0){10}}
\put(20,2){\circle*{1}}
\put(20,-2){\circle*{1}}
\put(30,2){\circle*{1}}
\put(30,-2){\circle*{1}}
\put(25,0){\circle{4}}
\put(35,0){\makebox(0,0){+}}
\put(40,0){\line(1,0){10}}
\put(50,0){\circle*{1}}
\put(40,0){\circle*{1}}
\put(45,2){\circle{4}}
\put(45,6){\circle{4}}
\put(55,0){\makebox(0,0){+}}
\put(60,0){\line(1,0){15}}
\put(60,0){\circle*{1}}
\put(75,0){\circle*{1}}
\put(67.5,0){\circle{6}}
\put(83,0){\makebox(0,0){$+\cdots$}}
\end{picture}
\label{a15}
\end{equation}
with the notation (\ref{a6}).
We note here that without using (\ref{a3}) to (\ref{a5})
we can directly obtain  (\ref{a10}), (\ref{a12}) and higher
orders if we note the equation corresponding to (\ref{PHI2}).
This point is taken in the following case of
$\langle\varphi(x)\varphi(y)\rangle$.
It is easy to convince oneself that if one uses
$(\Box+m^2+\lambda\phi^2/2)^{-1}$ instead of
$(\Box+m^2)^{-1}$ then the result of \cite{J} is obtained.
\subsubsection*{Case of $\langle\varphi(x)\varphi(y)\rangle$}
Now we consider the effective action of the bilocal composite
operator.
The generating functional $W[J]$ in this case is defined as
in (\ref{1}) with $S[\varphi,J]$ replaced by
\begin{eqnarray}
S[\varphi,J] &=& \!\!
- \frac{1}{2} \int \! d^{4} x
\varphi(x) (\Box+m^{2}) \varphi(x)
- \frac{\lambda}{4!} \int \!  d^{4} x  \varphi(x)^{4}
+ \frac{1}{2} \int \! d^{4}x d^4 y
J(x,y) \varphi(x)\varphi(y)
\nonumber \\
& \equiv & \!\!
- \frac{1}{2} \varphi G^{-1}_J \varphi
- \frac{\lambda}{4!} \varphi^{4},
\label{a17}
\end{eqnarray}
\begin{equation}
G^{-1}_J \equiv G^{-1}_J(x,y) =
(\Box+m^{2})\delta^{4}(x-y)-J(x,y).
\end{equation}
Note here that $J(x,y)$ has been absorbed in the
propagator $G_J$.
We define $\phi(x,y)$ and $\Gamma[\phi]$ by
\begin{equation}
\phi(x,y)=\frac{\delta W}{\delta J(x,y)}
\equiv \frac{1}{2}\langle \varphi(x)\varphi(y) \rangle,
\label{a18}
\end{equation}
\begin{equation}
\Gamma[\phi]=W[J]- \int d^{4} x d^4 y J(x,y) \phi(x,y).
\label{a19}
\end{equation}
Then the $0$-th order inversion formula (\ref{16}),
which is directly obtained as the $0$-th
order of  the relation corresponding to (\ref{PHI2}), gives
(with the space-time index omitted)
\begin{equation}
\phi=
\setlength{\unitlength}{1mm}
\begin{picture}(15,8)(-1,-1)
\thicklines
\put(0,0){\line(1,0){10}}
\end{picture}.
\label{a23}
\end{equation}
The line denotes the propagator $G$
evaluated at $J=J^{(0)}$, namely,
\begin{equation}
\phi=\frac{1}{i}\frac{1}{\Box+m^2-J^{(0)}}\equiv
\frac{1}{i}G^{(0)}.
\label{a24}
\end{equation}
Unlike the local case the key point is that this relation
can be explicitly inverted to give $J^{(0)}$
(compare with (\ref{0th})), that is,
\begin{equation}
J^{(0)}=\Box+m^2+i\phi^{-1},
\label{a25}
\end{equation}
which gives, by integration,
\begin{equation}
\Gamma^{(0)}={\rm Tr}\left(\Box+m^2\right)\phi+i{\rm Tr}\ln\phi.
\label{a26}
\end{equation}
Eq.~(\ref{17}) or the inversion formula of order $\lambda$,
which is obtained by the first order of the equation
like (\ref{PHI2}), gives
\begin{equation}
\setlength{\unitlength}{1mm}
\begin{picture}(15,8)(-1,-1)
\thicklines
\put(0,0){\line(1,0){15}}
\put(7.5,0){\circle*{1}}
\put(7.5,3){\makebox(0,0){$J^{(1)}$}}
\put(20,0){\makebox(0,0){+}}
\put(25,0){\line(1,0){10}}
\put(30,2){\circle{4}}
\put(40,0){\makebox(0,0){$=0$}}
\end{picture}
\label{a27}
\end{equation}
or, through integration,
\begin{equation}
\Gamma^{(1)}=
\setlength{\unitlength}{1mm}
\begin{picture}(18,8)(-1,-1)
\thicklines
\put(3,0){\circle{6}}
\put(8.8,0){\circle{6}}
\end{picture}.
\label{a28}
\end{equation}
We have used the notation in which
\setlength{\unitlength}{1 mm}
\begin{picture}(25,10)
\thicklines
\put(0,0){\line(1,0){20}}
\put(10,0){\circle*{1}}
\put(10,1){\makebox(0,0)[b]{$J^{(1)}$}}
\end{picture}
stands for
$G^{(0)}_{xz}J^{(1)}_{zw}G^{(0)}_{wy}$ where $\frac{1}{i}G^{(0)}=\phi$
(see (\ref{a24})).
 From (\ref{a28}) we make sure that $\Gamma^{(1)}$ is
actually a functional of the {\em bilocal} variable $\phi$
because lines in the graphs represent $\phi$.
The second order formula (\ref{18}) given by the
equation like (\ref{PHI2}) is written as,
\begin{eqnarray}
\lefteqn{
\setlength{\unitlength}{1mm}
\begin{picture}(15,13)(-1,-1)
\thicklines
\put(0,0){\line(1,0){15}}
\put(7.5,0){\circle*{1}}
\put(7.5,3){\makebox(0,0){$J^{(2)}$}}
\put(20,0){\makebox(0,0){+}}
\put(25,0){\line(1,0){25}}
\put(33,0){\circle*{1}}
\put(42,0){\circle*{1}}
\put(33,3){\makebox(0,0){$J^{(1)}$}}
\put(42,3){\makebox(0,0){$J^{(1)}$}}
\put(55,0){\makebox(0,0){+}}
\put(60,0){\line(1,0){25}}
\put(68,0){\circle*{1}}
\put(68,3){\makebox(0,0){$J^{(1)}$}}
\put(77,2){\circle{4}}
\put(90,0){\makebox(0,0){+}}
\put(95,0){\line(1,0){20}}
\put(105,2){\circle{4}}
\put(105,4){\circle*{1}}
\put(105,7){\makebox(0,0){$J^{(1)}$}}
\end{picture}
} \nonumber
\\
&+&
\setlength{\unitlength}{1mm}
\begin{picture}(15,13)(-1,-1)
\thicklines
\put(0,0){\line(1,0){25}}
\put(8,2){\circle{4}}
\put(17,0){\circle*{1}}
\put(17,3){\makebox(0,0){$J^{(1)}$}}
\put(30,0){\makebox(0,0){+}}
\put(35,0){\line(1,0){15}}
\put(40,2){\circle{4}}
\put(45,2){\circle{4}}
\put(55,0){\makebox(0,0){+}}
\put(60,0){\line(1,0){10}}
\put(65,2){\circle{4}}
\put(65,6){\circle{4}}
\put(75,0){\makebox(0,0){+}}
\put(80,0){\line(1,0){15}}
\put(87.5,0){\circle{6}}
\put(100,0){\makebox(0,0){$ = 0 .$}}
\end{picture}
\label{a29}
\end{eqnarray}
Using (\ref{a27}) we see that the 1 or 2-particle-reducible (2PR)
graphs in (\ref{a29}) exactly cancel out to give
\begin{equation}
J^{(2)}(x,y)=
\setlength{\unitlength}{1mm}
\begin{picture}(15,8)(-1,-1)
\thicklines
\put(6,0){\line(1,0){8}}
\put(10,0){\circle{8}}
\put(3,0){\makebox(0,0){$x$}}
\put(17,0){\makebox(0,0){$y$}}
\end{picture}
\end{equation}
or
\begin{equation}
\Gamma^{(2)}(x,y)=
\setlength{\unitlength}{1mm}
\thicklines
\begin{picture}(16,8)(-1,-1)
\thicklines
\put(5,0){\circle{6}}
\put(8,0){\circle{6}}
\end{picture}.
\end{equation}
As in the case of $\langle \varphi(x)\rangle$
we can continue the process and get
the well-known result;
\begin{equation}
\Gamma={\rm Tr}\left(\Box+m^2\right)\phi+
i{\rm Tr}\ln\phi+
{\cal K}_{\rm 2PI}[\phi]
\end{equation}
where
${\cal K}_{\rm 2PI}[\phi]$
is the original 2PI graph
${\cal K}_{\rm 2PI}[\frac{1}{i}\frac{1}{\Box+m^2}]$
with $\frac{1}{i}\frac{1}{\Box+m^2}$ replaced by $\phi$ or
\begin{equation}
{\cal K}_{\rm 2PI}[\phi]=
\setlength{\unitlength}{1mm}
\begin{picture}(15,8)(-1,-1)
\thicklines
\put(5,0){\circle{6}}
\put(10.75,0){\circle{6}}
\put(20,0){\makebox(0,0){$+$}}
\put(28,0){\circle{6}}
\put(31,0){\circle{6}}
\put(43,0){\makebox(0,0){$+\cdots$.}}
\end{picture}
\end{equation}

\setcounter{equation}{0}
\renewcommand{\theequation}{D.\arabic{equation}}
\section*{Appendix D --- Path-integral formula for the fermion
coherent state}
In this appendix we derive
(\ref{3.14})
from
(\ref{3.1}).
In order to clarify the notations, we first enumerate
some formulae for the fermionic coherent state in
the case of a single mode.
The generalization to multi-mode is straightforward.
For the anti-commuting operator
$a$, $a^{\dagger}$ like (\ref{3.6}), the coherent state
is defined as
\begin{equation}
a|z\rangle =z|z\rangle \quad,
\quad \langle z |a^{\dagger} =\langle z | z^*
\label{da1}
\end{equation}
where $z$ and $z^*$ are Grassmann numbers.
Then inner product of the two states becomes
\begin{equation}
\label{da2}
\langle z | z' \rangle = e^{z^* z'},
\end{equation}
which means that the coherent state is
neither normalized nor orthogonalized.
The matrix element in the coherent state is
\begin{equation}
\label{da3}
\langle z | {\cal O} ( a^{\dagger}, a ) | z' \rangle
=
{\cal O} (z^*, z') e^{z^* z'}
\end{equation}
where {\cal O} is a {\em normal-ordered}\/ operator.
The over-completeness is expressed as
\begin{equation}
\label{da4}
\int dz^* dz e^{-z^* z}|z\rangle \langle z |=1.
\end{equation}
The trace of a normal-ordered operator becomes
\begin{equation}
\label{da5}
{\rm Tr} {\cal O}(a^\dagger, a)=
\int dz^* dz e^{-z^* z}
\langle -z | {\cal O} ( a^\dagger, a ) | z \rangle.
\end{equation}
In order to derive (\ref{3.14}),
we first estimate
\begin{equation}
\langle z_F | T_\tau
e^{-\int d\tau (t_{\alpha \beta} a^\dagger _{\alpha}
a_{\beta}+ {\cal V}(a^\dagger_{\gamma} a_{\gamma}))} |
z_{I} \rangle e^{-z^*_{F \alpha} z^{ }_{F \alpha}}.
\label{da6}
\end{equation}
Here $\cal V$ is the on-site Coulomb term and
the source term appearing in (\ref{3.3}) to (\ref{3.5})
and
${\cal V}(a^\dagger_{\gamma} a_{\gamma}),$ $z_I$ and $z_F$
are abbreviations of
${\cal V}(\{a^\dagger_{\gamma}\},  \{a_{\gamma}\} ),$
$\{z_{I \gamma} \}$
and
$\{z_{F \gamma} \}$ respectively.
As usual we divide the exponential into $N+1$ pieces and insert
$N$ multi-mode complete sets like (\ref{da4}).
We get
\begin{equation}
\label{da7}
\left( \prod_{i=1}^{N} \prod_{\alpha}
\int dz_{i \alpha}^* dz^{ }_{i \alpha} \right)
e^{- \sum_{i=1}^{N+1} z^*_{i \alpha} z^{ }_{i \alpha}}
e^{- \sum_{i=1}^{N+1} z^*_{i \alpha} z^{ }_{i-1 \alpha}}
e^{-\varepsilon \sum_{i=1}^{N+1}
\{ t^{ }_{\alpha \beta} z^*_{i \alpha} z^{ }_{i-1 \beta}
+
{\cal V}(z^*_{i \gamma}  z^{ }_{i-1 \gamma} )\}}
\end{equation}
where $\varepsilon =\beta /(N+1)$,
$z_{0 \alpha}=z_{I \alpha}$,
$z_{N+1 \alpha}=z_{F \alpha}$
and we have assumed $\cal V$ is normal-ordered.
The first two exponential can be
{\em formally} written as
\begin{equation}
\label{da8}
e^{- \varepsilon \sum_{i=1}^{N+1} z^*_{i \alpha} (
 z^{ }_{i \alpha}
 -z^{ }_{i-1 \alpha})/\varepsilon}
\rightarrow
e^{-\int_0^{\beta} d \tau
z^*_{\alpha} (\tau)
\dot{z}^{ }_{\alpha} (\tau)}.
\end{equation}
In this way, through the trace formula (\ref{da5}),
we obtain the path-integral representation of
(\ref{3.1}), arriving at (\ref{3.14}).

\newpage
\setlength{\unitlength}{1pt}
\begin{picture}(24,10)(0,-12)
\thicklines
\put(12,0){\circle{24}}
\put(24,0){\circle*{3}}
\put(25,0){\makebox(0,0)[l]{$J^{(i)}$}}
\end{picture}
\\
Fig.1: The self-contraction of the pseudo-vertex.
\\
\vspace*{16pt}
\\
\vspace*{24pt}
\includegraphics{f1.ps}
\\
Fig.2: The self-contraction of the 4-point-vertex.
\\
\vspace*{16pt}
\\
\begin{picture}(24,10)(0,-12)
\begin{picture}(72,10)
\thicklines
\put(12,0){\circle{24}}
\put(24,0){\circle{24}}
\put(48,0){\circle{24}}
\put(60,0){\circle{24}}
\end{picture}
\hspace*{24pt}
\begin{picture}(74,10)
\thicklines
\put(12,0){\makebox(0,0)[r]{\footnotesize $J^{(2)}$}}
\put(26,0){\circle{24}}
\put(14,0){\circle*{3}}
\put(50,0){\circle{24}}
\put(62,0){\circle{24}}
\end{picture}
\hspace*{24pt}
\begin{picture}(74,24)
\thicklines
\put(12,0){\makebox(0,0)[r]{\footnotesize $J^{(2)}$}}
\put(26,0){\circle{24}}
\put(14,0){\circle*{3}}
\put(50,0){\circle{24}}
\put(62,0){\circle*{3}}
\put(64,0){\makebox(0,0)[l]{\footnotesize $J^{(2)}$}}
\end{picture}
\end{picture}
\\
\hspace*{31pt}(a)\hspace*{90pt}(b)\hspace*{84pt}(c)
\\
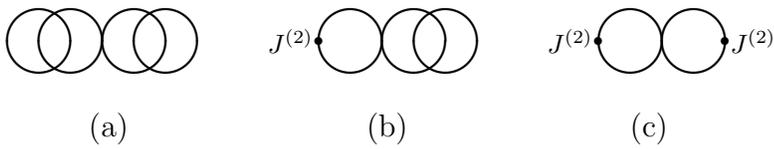
Fig.3: An example of 1VR graph that is canceled in
$\Gamma^{(5)}$.
\\
\vspace*{16pt}
\\
\begin{picture}(24,10)(0,-12)
\begin{picture}(120,10)
\thicklines
\put(12,0){\circle{24}}
\put(24,0){\circle{24}}
\put(48,0){\circle{24}}
\put(72,0){\circle{24}}
\put(96,0){\circle{24}}
\put(108,0){\circle{24}}
\end{picture}
\begin{picture}(32,10)
\put(15,0){\makebox(0,0){$\Rightarrow$}}
\end{picture}
\begin{picture}(48,10)
\thicklines
\put(12,0){\circle{24}}
\put(24,0){\circle{24}}
\put(36,0){\circle*{3}}
\end{picture}
\begin{picture}(35,10)
\thicklines
\put(12,0){\circle{24}}
\put(0,0){\circle*{3}}
\put(24,0){\circle*{3}}
\end{picture}
\begin{picture}(35,10)
\thicklines
\put(12,0){\circle{24}}
\put(0,0){\circle*{3}}
\put(24,0){\circle*{3}}
\end{picture}
\begin{picture}(48,10)
\thicklines
\put(12,0){\circle{24}}
\put(24,0){\circle{24}}
\put(0,0){\circle*{3}}
\end{picture}
\end{picture}
\\
Fig.4: An example of the graph of $N(\bar{\cal K})=4$.
\\
\vspace*{16pt}
\\
\vspace*{14pt}
\includegraphics{f2.ps}
\\
Fig.5: The 1-part (encircled by the dashed line)
in a graph of $\phi$.
\\
\vspace*{100pt}
\includegraphics{f3.ps}
\\
Fig.6: The procedure to reach the second largest
structure.\\
\hspace*{20pt}(a) proceed to a larger 1-part.\\
\hspace*{20pt}(b) reach the second largest 1-part.\\
\hspace*{20pt}(c) reach the second largest 1-parts.\\
\hspace*{20pt}(d) reach the second largest (1-part) structure.
\end{document}